\documentclass{aa}
\usepackage{natbib}
\usepackage{amsmath}
\usepackage{gensymb}
\usepackage{placeins}
\usepackage{afterpage}
\usepackage{tabularx}
\usepackage{booktabs}
\usepackage{multirow}
\usepackage{kantlipsum}

\usepackage{color}
\usepackage{graphicx}
\usepackage{txfonts}
\usepackage{twoopt}
\usepackage{hyperref}
\usepackage{etoolbox}
\makeatletter
\newcount\c@additionalboxlevel
\newcount\c@maxboxlevel
\patchcmd\@combinedblfloats{\box\@outputbox}{%
  \stepcounter{additionalboxlevel}%
  \box\@outputbox
}{}{\errmessage{\noexpand\@combinedblfloats could not be patched}}

\AtBeginShipout{%
  \ifnum\value{additionalboxlevel}>\value{maxboxlevel}%
    \typeout{Warning: maxboxlevel might be too small, increase to %
      \the\value{additionalboxlevel}%
    }%
  \fi 
  \@whilenum\value{additionalboxlevel}<\value{maxboxlevel}\do{%
    \typeout{* Additional boxing of page `\thepage'}%
    \setbox\AtBeginShipoutBox=\hbox{\copy\AtBeginShipoutBox}%
    \stepcounter{additionalboxlevel}%
  }%
  \setcounter{additionalboxlevel}{0}%
}
\makeatother

\bibpunct{(}{)}{;}{a}{}{,} 
\makeatletter
\newcommandtwoopt{\citeads}[3][][]{\href{http://adsabs.harvard.edu/abs/#3}%
{\def\hyper@linkstart##1##2{}%
\let\hyper@linkend\@empty\cite[#1][#2]{#3}}}
\newcommandtwoopt{\citealpads}[3][][]{\href{http://adsabs.harvard.edu/abs/#3}%
{\def\hyper@linkstart##1##2{}%
\let\hyper@linkend\@empty\citealp[#1][#2]{#3}}}
\newcommandtwoopt{\citepads}[3][][]{\href{http://adsabs.harvard.edu/abs/#3}%
{\def\hyper@linkstart##1##2{}%
\let\hyper@linkend\@empty\citep[#1][#2]{#3}}}
\newcommandtwoopt{\citetads}[3][][]{\href{http://adsabs.harvard.edu/abs/#3}%
{\def\hyper@linkstart##1##2{}%
\let\hyper@linkend\@empty\citet[#1][#2]{#3}}}
\newcommandtwoopt{\citeyearads}[3][][]%
{\href{http://adsabs.harvard.edu/abs/#3}
{\def\hyper@linkstart##1##2{}%
\let\hyper@linkend\@empty\citeyear[#1][#2]{#3}}}
\makeatother

\newcommand{\cng}[1]{{#1}\color{black}} 

\title{{Dust properties across the CO snowline in the HD~163296 disk from ALMA and VLA observations}}

\author{G. Guidi\inst{1,2}\and M. Tazzari\inst{3,4}\and L. Testi\inst{1,3,4}\and  I. de Gregorio-Monsalvo\inst{3,5}\and C.~J.~Chandler\inst{6}\and L.~P\'erez\inst{6,7}\and A.~Isella\inst{8}\and A.~Natta\inst{1,9}\and S.~Ortolani\inst{10,11}\and Th.~Henning\inst{12}\and S.~Corder \inst{5} \and  H.~Linz \inst{12} \and S.~Andrews \inst{13} \and D.~Wilner \inst{13} \and L.~Ricci \inst{13} \and J.~Carpenter \inst{14} \and A.~Sargent \inst{14} \and L.~Mundy \inst{15} \and S.~Storm \inst{15} \and N.~Calvet \inst{16} \and C.~Dullemond \inst{17} \and  
J.~Greaves \inst{18} \and J.~Lazio \inst{19} \and A.~Deller \inst{20} \and W.~Kwon \inst{21}  }

\institute{
        INAF-Osservatorio Astrofisico di Arcetri, Largo E. Fermi 5, I-50125 Firenze, Italy 
         \and
                Dipartimento di Fisica e Astronomia, Universit\`a degli Studi di Firenze, Italy 
         \and
              ESO, Karl Schwarzschild str. 2, D-85748 Garching bei M\"unchen, Germany
         \and   
              Excellence Cluster `Universe', Boltzmannstr. 2, D-85748 Garching bei Muenchen, Germany
              \and 
              Joint ALMA Observatory (JAO), Alonso de Cordova 3107 Vitacura -Santiago de Chile
              \and 
             National Radio Astronomy Observatory, Socorro NM 87801, USA 
             \and
             Jansky Fellow
             \and
			Department of Physics and Astronomy, Rice University, 6100 Main Street, Houston, TX, 77005, USA
            \and
            School of Cosmic Physics, Dublin Institute for Advanced Studies, 31 Fitzwilliams Place, 2 Dublin, Ireland
            \and
            Dipartimento di Fisica e Astronomia, Universit\`a di Padova, Italy 
            \and
            INAF-Osservatorio Astronomico di Padova, Vicolo dell’Osservatorio 5, I-35122 Padova, Italy
            \and
            Max Planck Institut f\"ur Astronomie, K\"onigstuhl 17, D-69117 Heidelberg, Germany
            \and 
            Harvard-Smithsonian Center for Astrophysics, 60 Garden Street, Cambridge, MA 02138, USA
            \and
            Department of Astronomy, California Institute of Technology, MC 249-17, Pasadena, CA 91125, USA
            \and
            Department of Astronomy, University of Maryland, College Park, MD 20742
            \and
            Department of Astronomy, University of Michigan, 830 Dennison Building, 500 Church Street, Ann Arbor, MI 48109, USA
            \and 
           Institute for Theoretical Astrophysics, Heidelberg University, Albert-Ueberle-Strasse 2, D-69120 Heidelberg, Germany
           \and 
            SUPA, School of Physics and Astronomy, University of St Andrews, North Haugh, St Andrews,  KY16 9SS Scotland, UK
            \and 
            Jet Propulsion Laboratory, California Institute of Technology Pasadena, CA 91109, USA
            \and
            The Netherlands Institute for Radio Astronomy (ASTRON), Dwingeloo, The Netherlands
            \and
            Korea Astronomy and Space Science Institute, 776 Daedeok-daero, Yuseong-gu, Daejeon 34055, Republic of Korea
}

\begin{document}
\titlerunning{Dust properties across the CO snowline in HD163296}
\authorrunning {Guidi et al.}

\abstract
{To characterize the mechanisms of planet formation it is crucial to investigate the properties and evolution of protoplanetary disks around young stars, where the initial conditions for the growth of planets are set.
The high spatial resolution of Atacama Large Millimeter/submillimeter Array (ALMA) and Karl G. Jansky Very Large Array (VLA) observations allows now the study of radial variations of dust properties in nearby resolved disks and the investigatation of the early stages of grain growth in disk midplanes.} 
{Our goal is to study grain growth in the well-studied disk of the young, intermediate mass star HD~163296 where dust processing has already been observed, and to look for evidence of growth by ice condensation across the CO snowline, already identified in this disk with ALMA.}
{Under the hypothesis of optically thin emission we compare images at different wavelengths from ALMA and VLA to measure the opacity spectral index across the disk and thus the maximum grain size. We also use a Bayesian tool based on a two-layer disk model to fit the observations and constrain the dust surface density.}
{The measurements of the opacity spectral index indicate 
the presence of large grains and pebbles ($\geq$1~cm) in the inner regions of the disk (inside $\sim$50~AU) and smaller grains, consistent with ISM sizes, in the outer disk (beyond 150~AU).
Re-analysing ALMA Band 7 Science Verification data we find (radially) unresolved excess continuum emission centered near the location of the CO snowline at $\sim$90 AU.}
{Our analysis suggests a grain size distribution consistent with an enhanced production of large grains at the CO snowline and consequent transport to the inner regions. Our results combined with the excess in infrared scattered light found by Garufi et al. (2014) 
suggests the presence of a structure at 90~AU involving the whole vertical extent of the disk. This could be evidence for small scale processing of dust at the CO snowline.}
\keywords{protoplanetary disks -- stars: circumstellar matter -- stars: formation -- radio continuum: planetary systems} 
\maketitle

\begin{figure*}[ht!]
 \centering
 \includegraphics[width=16cm]{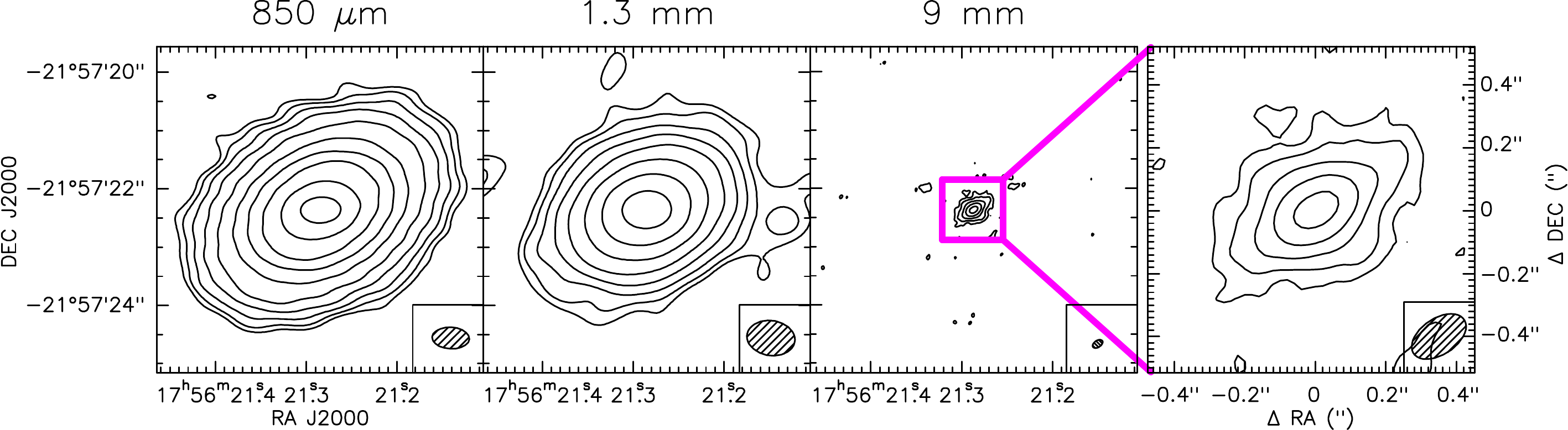}

 \caption{Continuum maps at different wavelengths, from the left: 850~$\mu$m, 1.3~mm and 9~mm, this latter obtained from the combination of VLA 8.0~mm and 9.8~mm bands. The contour levels 
 correspond to -3 (dashed), 3, 6, 12, 24, 48, 100, 200, 400, 800, 1600 $\sigma$. The bottom right of every panel shows the synthesized beam (see Table~\ref{table:param}).}
 \label{fig:cont}
\end{figure*}
\begin{table*}[!th]
\centering
\begin{tabular*}{0.9\textwidth}{l @{\extracolsep{\fill}} cccccccc}
\hline \hline \\
 & $\lambda$ & $\nu$& $F_{int}$&$F_{peak}$ & rms &CLEAN beam &Beam P.A.\\
 &\small{[mm]}&\small[GHz] &\small[Jy]&\small[Jy/beam]& \small[mJy/beam]&\small[FWHM]& [$^\circ$]\\
 \midrule
ALMA B7& 0.85 	& 352.9 & 2.13 $\pm$ 0.21 & 
0.44 & 0.18 &$0.57''\times0.37''$ & \phantom{0}86.5\\
ALMA B6& 1.33& 225.3& 0.60 $\pm$ 0.06 & 0.22 
&0.37 &$0.74''\times0.60''$ & \phantom{0}77.3 \\
\vspace{2mm}
VLA Ka & 9.00 & 34.0 & (1.83 $\pm$ 0.02) $\cdot 10^{-3}$ &0.847$\cdot 10^{-3}$ & 0.013 &$0.18''\times0.11''$ & -49.4\\ 
VLA Ka&8.00&\phantom{0}37.5&(2.02 $\pm$ 0.2) $\cdot 10^{-3}$ &0.928 $\cdot 10^{-3}$& 0.024 &$0.16''\times0.10''$ & $-$56.1\\
VLA Ka&9.83&\phantom{0}30.5&(1.65 $\pm$ 0.2) $\cdot 10^{-3}$ &0.806 $\cdot 10^{-3}$& 0.015 &$0.19''\times0.12''$ & \phantom{0}131.5\\
\bottomrule
\end{tabular*} 
\caption{Parameters for the deconvolved images displayed in Fig.~\ref{fig:cont} and for the images obtained from the single VLA frequencies of 30 and 37~GHz.}
\label{table:param}
\end{table*}

\section{Introduction}
Understanding the mechanisms that cause small dust particles to grow into larger bodies and eventually form planets is still an open issue 
in spite of the large laboratory, theoretical and observational efforts  \citepads[see the review by][]{Testi2014}.
These processes are thought to occur in circumstellar disks around young stars on rather short timescales: infrared surveys show that the fraction of low mass stars with disks drops 
dramatically for stars older than 10 Myr \citepads{mamaj,hern}, suggesting that in order to form large solids before the disk disperses the growth process has to take place within a few million years. 
Analysis of dust properties in protostellar envelopes and young circumstellar disks suggests 
the formation of large grains 
(even up to few millimeter size) 
during the disk formation stage in infalling envelopes \cng{\citepads{kwon2009,Miotello2014}}; 
still, substantial growth of dust grains occurs  
in the cold midplane of protoplanetary disks, where most of the solid mass is confined and 
planetesimals and planets are thought to form. The recent ALMA results for the young protoplanetary disk surrounding HL~Tau suggest that planet formation may indeed occur very early in the disk lifetime \citepads{ALMAHLTau,dipierro2015, tamayo2015}. 

The life of a growing dust grain in a protoplanetary disk is not an easy one. It was realized already four decades ago that aerodynamical friction may effectively prevent grain growth \citepads{Weidenschilling77}. The process of rapid radial migration and fragmentation effectively sets an upper limit to the grain sizes as a function of radius and a very rapid evolutionary timescale for dust particles in disks \citepads{Brauer2008,Birnstiel2012}. These theoretical expectations are at odds with direct observations of dust properties in the outer disks from millimeter observations \citepads{Ricci2010,Birnstiel2010}. This is a general result, although some authors have shown that disks with peculiar growth processes \citepads[e.g.,][]{Laibe2014,drazk2014} or specific dust properties \citepads{okuzumi2012} may retain large particles more efficiently. To overcome the general inconsistency between models and observations, the most commonly accepted scenarios involve local grain growth and trapping in  small regions, with sizes of the order or smaller than the local disk scale height \citepads{klahr97,Pinilla2012,Testi2014,Johansen2014}.

The regions in the disk midplane that correspond to the snowlines of major volatiles are particularly interesting. These may promote locally efficient grain growth through re-condensation across the snowline or by changing the sticking properties of ice-coated grains and, in addition, the local release of volatiles from the ices may induce a local pressure bump that could trap large grains \citepads[e.g.][]{2000Icar..146..525S,2009ApJ...702.1490W,RosJohansen2013,2015ApJ...798...34G}.

The dense regions of disk midplanes can be investigated at sub-millimeter and millimeter wavelengths, 
where the dust emission is more optically thin and can now be spatially resolved with present-day facilities, like the Atacama Large Millimeter/submillimeter 
Array (ALMA) 
and the Karl G. Jansky Very Large Array (VLA), among others. 
These facilities allow us to obtain sensitive and high angular resolution observations of the dust continuum and gas emission lines from protoplanetary disks  
and therefore study the radial variations of the physical parameters of the disk.
Recent results include the study of the dust properties as a function of radius in disks \citepads[e.g.,]{Guilloteau2011,Banzatti2011,2012ApJ...760L..17P,Perez15,menu2014,Tazzari2015},
and the clear identification of the CO snowline in nearby disks \citepads[][]{Mathews2013,Qi2013,Qi2015}, through its effects on DCO$^+$ and N$_2$H$^+$ abundances. It is thus becoming possible to directly investigate observationally the effect of the CO snowline on grain growth. This snowline is thought to be one of the most important after that of water, as the CO abundance in protostellar ices is found to be about 30-40\%\ of H$_2$O \citepads{Oeberg2011}. 
In this work we focus on the Herbig~Ae star HD~163296, a bright and isolated object at a distance of $122^{+17}_{-13}$ parsec \citepads{vdancker} with a relatively massive disk \citepads[$\sim$ 0.1$M_{\odot}$:][]{qi2011,isella} 
and an excellent prototype for gas- and dust-rich protoplanetary disks.
An estimated age of 5 Myr was obtained from the comparison between Hipparcos astrometric measurements and pre-main sequence evolutionary models by \citetads{vdancker}. The stellar parameters computed by \citetads{natta04} are $M_* = 2.3 M_{\odot}$, 
$L_* = 36 L_{\odot}$, $T_{eff}=9500K$. 
To study the dust properties in the protoplanetary disk around HD~163296, we re-analyse the ALMA Science Verification observations \citepads{itziar,Mathews2013,rosen} discussing for the first time the continuum emission in Band 6, combined with new VLA observations from the Disks@EVLA collaboration. In $\S 2$ we describe the observational data, in $\S 3$ we report the main new results of our analysis, in $\S 4$ we present the result of our disk modeling, and in $\S 5$ we discuss the main implications for grain properties. 

\section{Observations} 
\label{sec:obs}
\subsection{ALMA Observations}
The ALMA observations of HD~163296 (also known as MWC~275)  were part of the ALMA Science Verification Program 2011.0.000010.SV\footnote{The ALMA Science Verification data can be found at: {\tt https://almascience.eso.org/alma-data/science-verifica\\tion}}. 

Band 6 observations were performed on 2012 June 9, June 23, and July 7 using a set of configurations comprising 20, 21 and 19 antennas, respectively. 
The total integration time was 3.14 hours (1.4 hours on the science source), the field of view was $\sim$20 arcseconds and the baselines 
ranged from 20 to 400 meters, corresponding to spatial scales of 1600~AU to 80~AU at the distance of the object. The flux density calibrator for the three execution blocks were Juno, Neptune 
and Mars respectively, while the phase calibrator was 
J1733-130 and the bandpass calibrator was J1924-292. 
The correlator was set with four spectral windows in dual polarization mode, two spectral windows in the upper side band and two in the lower side band:
two spectral windows, $\#$0 (216.2 - 218 GHz) and $\#$3 (233.1 - 234.9 GHz), were used to observe the line-free continuum with channel widths of 488 kHz; 
while at higher resolution (244 kHz) spectral window $\#$1 (219.5 - 220.4 GHz) 
included the C$^{18}$O(2~$-$~1) line at 219.560~GHz and the $^{13}$CO(2~$-$~1) line 
at 220.398~GHz, and spectral window $\#$2 (230.5 - 231.5 GHz) covered the CO(2~$-$~1) line at 230.539~GHz. 
Imaging of the continuum emission of HD 163296 was performed excluding the above-mentioned lines \citepads[see][]{rosen,klaassen2013}. 
\cng{Data calibration was performed using version 4.1.0 of the Common Astronomy Software Application (CASA), self calibration was applied making use of 
the line-free channels}, and using robust weighting during the CLEAN deconvolution we were able to reach a resolution of $0.74"\times0.60"$ with a rms of 0.37 mJy/beam (see Table~\ref{table:param}).

Band 7 observations were made on 2012 June 9, June 11 and June 22, with the same antenna configurations as for Band~6. Collectively, the five datasets 
covered an integration time of 3.9 hours, with 2.3 hours on the science target. The flux density calibrators were Juno and Neptune, while the bandpass 
and phase calibrators were the same as those used for Band 6. The two spectral windows in the lower side band were $\#$2 (345.56 $-$ 346.03 GHz) and  $\#$3 
(346.52 $-$ 347.47) with channel widths of 122 and 244 kHz respectively; the ones in the upper side band were $\#$1 (356.50 $-$ 356.97 GHz) with 122 kHz channel width and $\#$0 (360.11 $-$ 360.23~GHz) at a high spectral resolution of 30.5 kHz; these included the emission lines: CO(3~$-$~2) at 345.796 GHz ($\#$2), 
 HCO$^{+}$(4 $-$ 3) at 356.734~GHz ($\#$1), H$^{13}$CO$^+$(4~$-$~3) at 346.998 GHz ($\#$3) and DCO$^+$(5~$-$~4) at 360.160 GHz ($\#$0). 
A detailed analysis of these spectral lines has been published by \citetads{Mathews2013}, \citetads{itziar}, and \citetads{rosen}. \cng{In this work we focus on the imaging of the continuum emission, obtained with the task CLEAN applying a robust weighting with Briggs parameter 0.5 and achieving a synthesized beam of $0.57"\times0.37"$ and a rms of 0.18 mJy/beam (see Table \ref{table:param}).}

\subsection{VLA Observations}
Observations of HD 163296 were made using the Karl G. Jansky Very Large Array (VLA) of the National Radio Astronomy Observatory\footnote{The National Radio Astronomy Observatory is a facility of the National Science Foundation operated under cooperative agreement by Associated Universities, Inc.} as part of the Disks@EVLA project (AC982) in 2011 May and June, in the BnA and A configurations. The Ka-band ($\lambda\sim$1cm) receivers were used with two 1GHz basebands centered at 30.5 and 37.5GHz, providing projected uv-spacings from 25 to 3,800~k$\lambda$.  The complex gain was tracked via frequent observations of J1755$-$2232, and the spectral shape of the complex bandpass was determined through observations of 3C279.  The absolute flux density scale was derived from observations of 3C286 \citepads[e.g.,]{perley2013}, and its overall accuracy is estimated to be 10\%. The data were calibrated, flagged, and imaged using a modified version of the VLA Calibration Pipeline (see https://science.nrao.edu/facilities/vla/data-processing/pipeline/scripted-pipeline) with CASA\@.  The astrometry reported here corresponds to that derived from the A configuration data.

In addition, HD 163296 was observed with the C-band ($\lambda\sim$6cm) receivers in the DnC configuration in September 2010, in order to evaluate any potential contamination from ionized gas at shorter wavelengths.  Two 1GHz basebands were centered at 5.3 and 6.3GHz.  Complex gain variations were tracked through observations of J1820$-$2528, and the bandpass and absolute flux density scale was obtained through observations of 3C286. The data were calibrated, flagged, and imaged using the CASA data reduction package.  HD 163296 was detected with integrated flux density $F_{\rm 5.2cm} = (410\pm 57)\mu$\mbox{Jy}.

\section{Observational results}
\label{sec:res}
\subsection{Continuum maps}
In Figure \ref{fig:cont} we show the continuum intensity maps obtained from the line-free channels at three different wavelengths: ALMA Band 7, ALMA Band 6 and the combination of the two VLA frequency ranges at 30.5 and 37.5 GHz. The parameters of the images are listed in Table \ref{table:param}. 

\begin{figure}
\centering
\includegraphics[keepaspectratio=true,width=10cm]{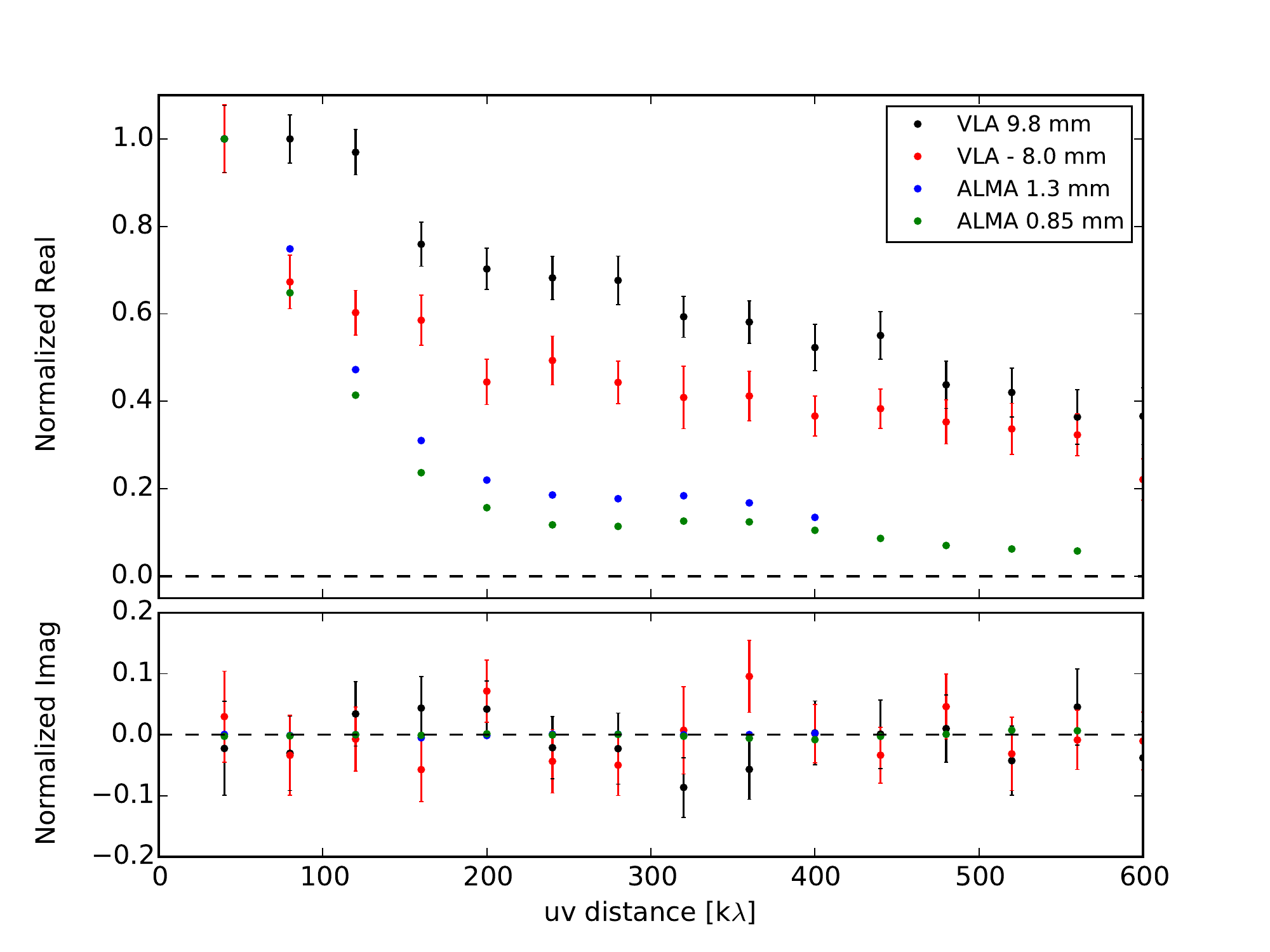}
\caption{Real and imaginary part of the measured visibilities as a function of uv-distance, deprojected assuming PA=133 \degree, {\it i}=44\degree and bin-averaged every 40 k$\lambda$. Visibilities at each wavelength have been normalized by the average value at 40~k$\lambda$ and error bars display the standard error of the mean, negligible for ALMA observations.}
\label{fig:vis}
\end{figure}

In agreement with previous work \citepads[][]{itziar,natta04} we observe more compact emission at longer wavelengths, obtaining at 850 $\mu$m a projected radius of $\sim$2.4$''$ at a 3 sigma level, corresponding to about 290~AU, while at 1.3~mm the outer radius of the emission is $\sim$260 AU, and is $\sim$40~AU at 8-10~mm. We note that the low signal-to-noise of the VLA data at 8 and 10~mm can lead to an underestimate of the extent of the emission at these wavelengths, and it is critical to consider the visibility function for a proper analysis of the disk structure. We use 44$\degree$ for the disk inclination and 133$\degree$ for the disk position angle \citepads[from][]{qi2011} and plot the normalized real and imaginary part of the visibilities in Figure \ref{fig:vis}.  These plots show that the real part of the visibilities decline more steeply at the shorter wavelengths (ALMA 850~$\mu$m and 1.3~mm) than at longer wavelengths (VLA 8~mm and 9.8~mm), demonstrating that the millimeter wavelength emission is intrinsically considerably more extended than the centimeter wavelength emission (a point source would be a constant 1.0 as a function of uv-distance in this plot). The integrated flux density above the $3\sigma$ level at 850 $\mu$m is $F_{850\mu m}=2.13\pm 0.02$~Jy, similar to the value found by \citetads{isella} and \citetads{itziar}, 
while in Band 6 we find $F_{1.3mm}=0.59\pm 0.06$~Jy. At the longer wavelengths the flux density decreases by almost 3 orders of magnitude, with $F_{8mm}=2.0\pm 0.2$~mJy and 
$F_{10mm}=1.6\pm 0.2$~mJy. We include a calibration error of 10\% in the measurements of the flux densities. 
We find at all wavelengths a smoothly decreasing intensity profile, consistent with the disk temperature and surface density decreasing with radius. 

\subsection{Proper motions}

\begin{figure}[ht!]
 \centering
 \includegraphics[width=9cm]{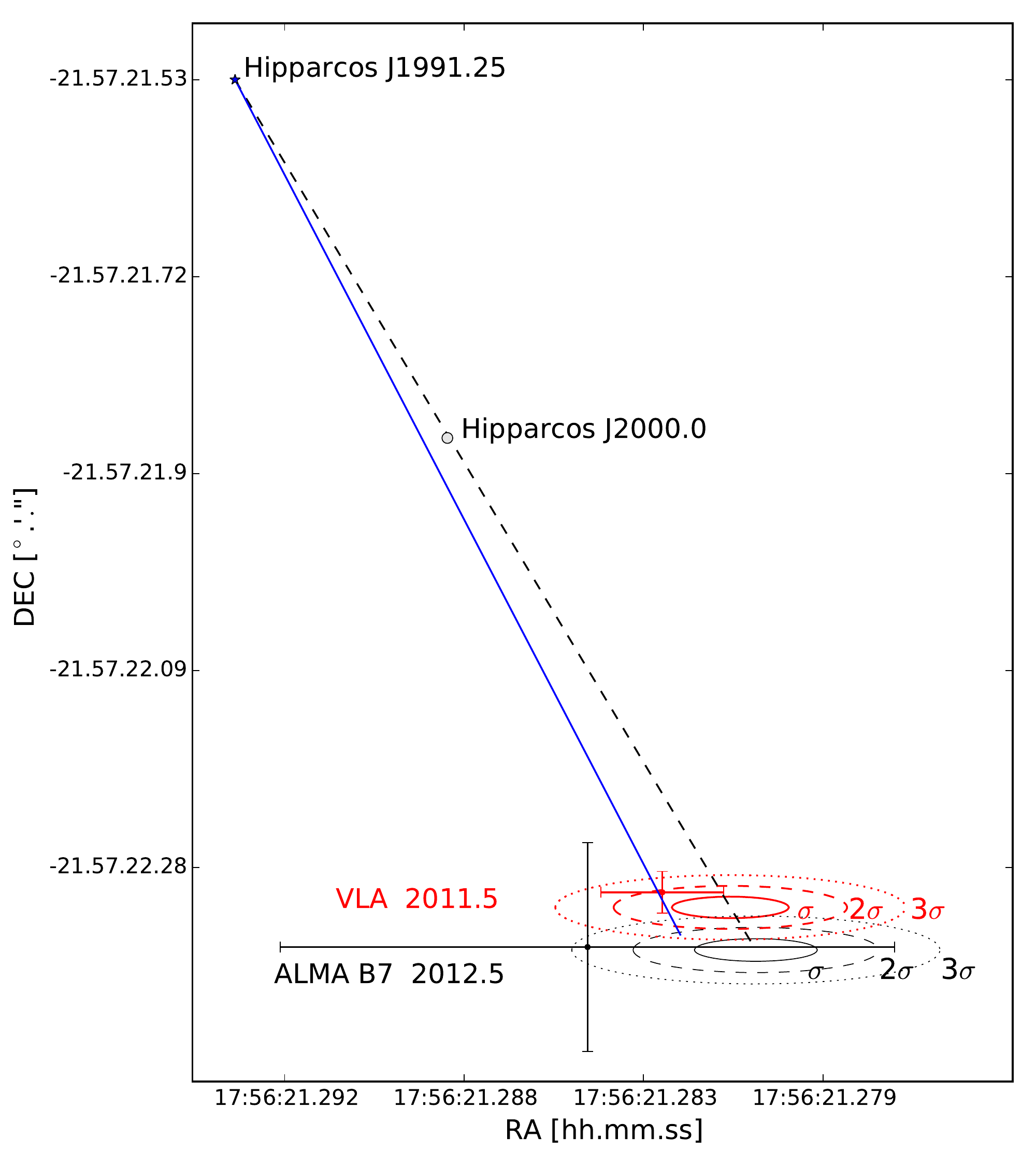}

 \caption{Position of HD 163296 at the different observing epochs. The dashed line represents the proper motions from the Hipparcos measurement (J1991.25) 
 to 2012, with respective proper motion error ellipses at 1, 2 and 3 $\sigma$ from the predicted positions. The black dot indicates our position estimate for the star at epoch 2012.5 based on the peak of the image in Band 7, while the red dot is the peak of the VLA image at 9~mm in 2011.5.  The error bars are given by the astrometric accuracy of 0.1 arcsec for ALMA and 0.02 arcsec for VLA.  The blue solid line represents the proper motions calculated from a least squares regression between Hipparcos measurements and the observations. }
\label{fig:prop}
\end{figure}

We checked whether the position of the star at the two different epochs of our observations was consistent with the proper motions reported in the literature: applying a gaussian 
fit with to the images we find that the positions of the peaks fall within the $3\sigma$ error ellipse from the predicted position based on the 
Hipparcos astrometric mission measurements (J1991.25). For the ALMA observations (2012) we used Band 7 data because of its better signal to noise, with the peak position of the images obtained from the calibrated dataset before the self-calibration was applied. t
In Figure \ref{fig:prop} are displayed the estimated positions of the central star; the astrometric error for interferometric observations is affected, among other things, by the phase calibration and depends on several factors (weather conditions, the separation between the target and the calibrator, etc.).  We assume here that the absolute astrometry of the ALMA data is $0.1''$\footnote{\tt https://help.almascience.org/index.php?/Knowledgebase\\/Article/View/153/6/what-is-the-astrometric-position-\\accuracy-of-an-alma-observation}; while for the VLA in A configuration it is expected to be $\sim0.02''$\footnote{\tt https://science.nrao.edu/facilities/vla/docs/manuals/\\oss/performance/positional-accuracy}. 
The proper motions derived from a least squares interpolation between our peaks at the two different epochs and the J1991.25 Hipparcos position are consistent 
with the latest reduction of the Hipparcos data \citepads{vanLeeuw} within the errors and are listed in Table \ref{table:proper}.
The main difference we find is in the Right Ascension, where our best fit would imply a smaller proper motion. Nevertheless, the difference is still well within the uncertainties. 

\begin{table}[]
 \begin{tabular*}{0.5\textwidth}{r@{\extracolsep{\stretch{1}}}cc}
 \toprule
 & RA $[hh:mm:ss]$ & DEC $[\degree .'.'']$ \\ 

 \midrule
 Hipparcos J1991.25 & 17:56:21.293 & -21.57.21.527 \\
 VLA 2011.5 & 17:56:21.283 & -21.57.22.30\phantom{0} \\
 ALMA 2012.5 & 17:56:21.285 & -21.57.22.36\phantom{0} \\
 \hline
 \end{tabular*}

 \begin{tabular*}{0.5\textwidth}{r@{\extracolsep{\stretch{1}}}*{3}{c}@{}}
 \hline
 &pm-ra & pm-dec & error ellipse\\
 &$[mas/yr]$ & $[mas/yr]$ & $[mas$ $mas]$\\
 \midrule
 Hipparcos & -7.98 & -39.21 & [0.94 0.51] \\
 this paper & -6.8 & -38.5 & [1.0 1.0]\\
 \bottomrule
 \end{tabular*}
 \vspace{3mm}
 \caption{Top table: right ascension and declination of HD 163296 at the different epochs from Hipparcos. Bottom table: proper motions with associated error ellipse from Hipparcos measurements and from the least squares interpolation performed in this paper.}
 \label{table:proper}
\end{table}

\subsection{SED and free-free contribution}
\label{sec:sed}
\begin{figure}[t!]
 \centering
 \includegraphics[keepaspectratio=true,width=10cm]{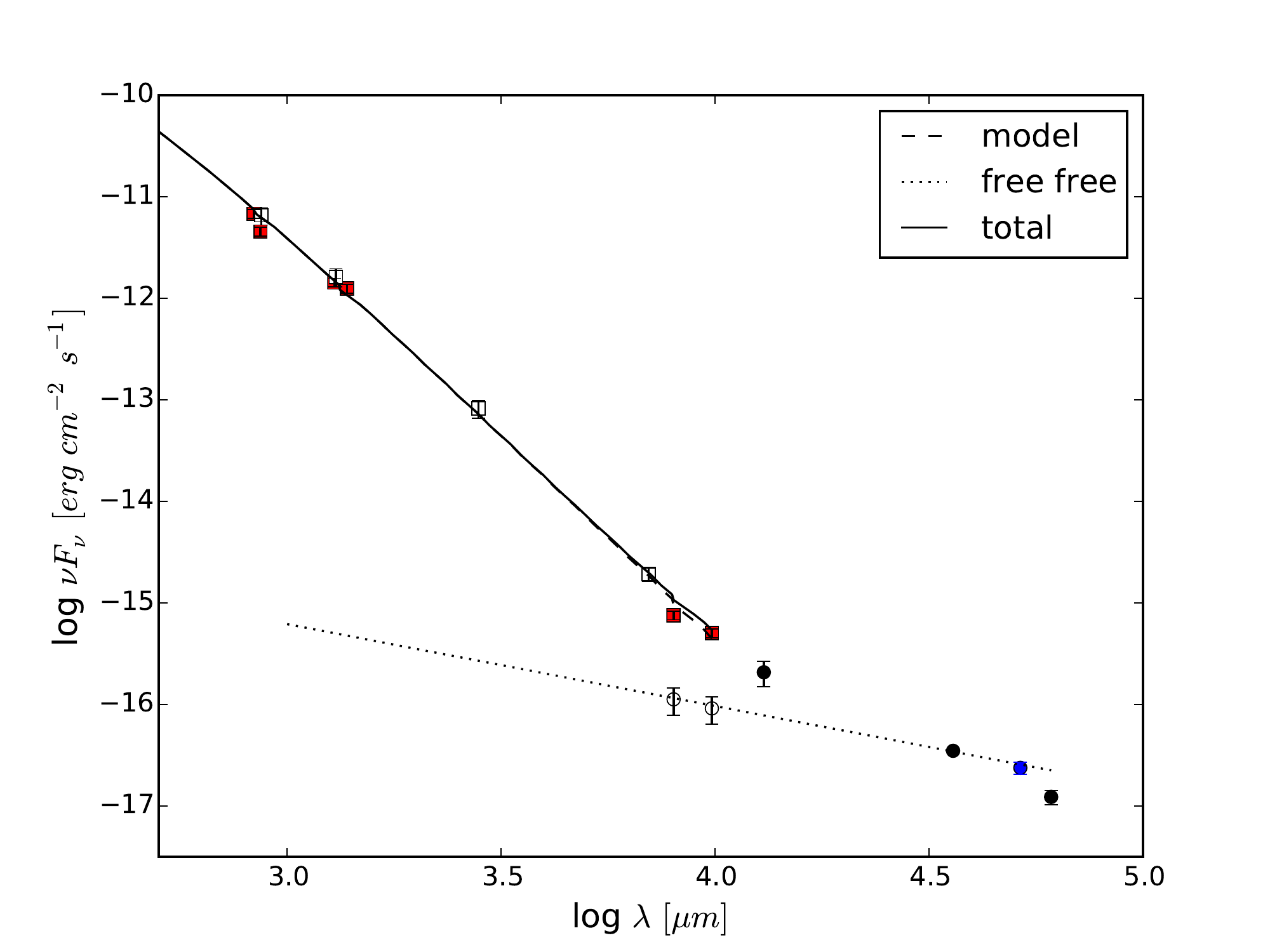}

 \caption{HD 163296 spectral energy distribution: our measurements are represented in red filled squares, the white empty squares are taken from the literature \citepads[][]{isella,natta04}, 
and the longer wavelength measurements used to evaluate the free free contribution are shown in full black circle marks \citepads{natta04} and blue full circle mark (Disks@EVLA collaboration).  
Empty circles show the value of 0.3 mJy for the free free emission estimated in this paper. 
The dashed curve shows the best fit model from this paper (see Section \ref{sec:modello}), the dotted line is the estimated free-free emission and the solid line the sum of the two.}
 \label{fig:sed}
\end{figure}

The integrated flux densities measured within a 3 sigma level in our observations and the spectral energy distribution predicted by our model (see Section \ref{sec:modello}) are plotted in Figure \ref{fig:sed}. 
Our model is consistent with the observed flux densities in the literature 
\citepads[][]{natta04,isella}.
At sub-millimeter and millimeter wavelengths the continuum emission is due to the dust in the colder regions of the disk midplane, while at centimeter wavelengths the emission may also arise from free electrons in the stellar wind. 
Since this contribution is thought to come from a region in the inner part of the disk, we 
examined the longest baselines ($\geq$1500~k$\lambda$) in the VLA 8.0 and 9.8~mm  observations and we estimate an upper limit of 0.3~mJy for this wind emission, corresponding to the asymptotic value reached at the higher uv-distances by the real part of the visibilities. 
We fit a power law for the free-free emission by performing a least squares interpolation between our estimates at 8 and 10~mm and other VLA measurements at 3.6~cm \citepads{natta04} and 5.2~cm (see Section \ref{sec:obs}). The resulting power law is $F_\nu\propto\nu^{-0.19\pm 0.11}$, and is shown with a dotted line in Figure \ref{fig:sed} (note that in figure we plot $\nu F_\nu$ as a function of $\lambda$). 
According to this estimate the free-free component at wavelengths shorter than 7~mm is negligible \citepads[see also][]{natta04}. 
 
\subsection{Excess emission at 850~$\mu$m}
\label{subsec:excess.emission}
\begin{figure*}[!ht]
\begin{center}
\includegraphics[keepaspectratio=true,width=15cm]{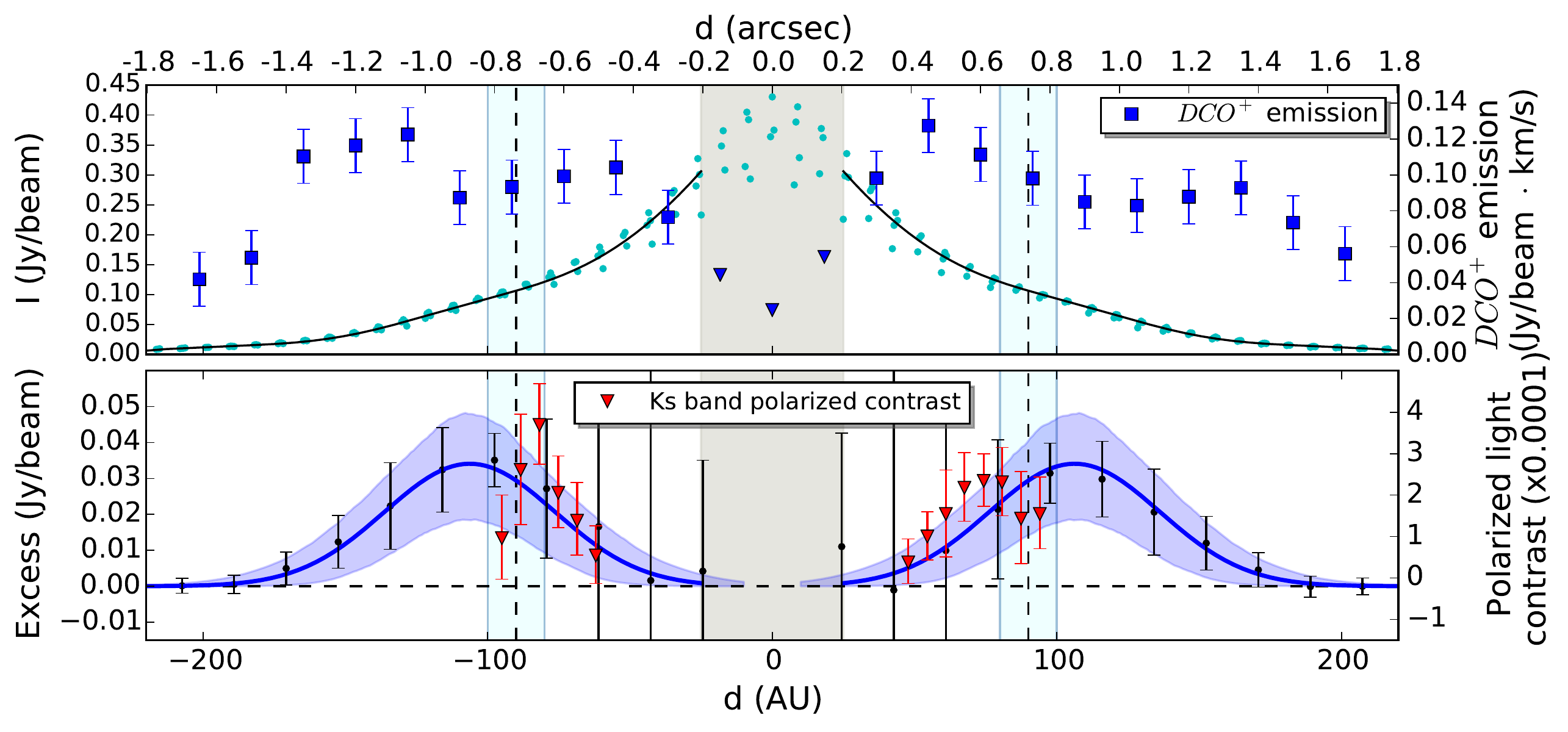}
\caption{\textit{Top panel}: flux density at 850~$\mu$m along the disk major axis from SE (left) to NW (right). The solid line represents the fit (polynomial profile + a gaussian for the excess) performed excluding the inner 0.2 arcsec of the disk (grey shadowed area). The vertical dashed line corresponds to the CO snowline at 90 $\pm$ 10~AU (from \citetads{Qi2015}). DCO$^+$ emission in blue squares binned by 0.15 arcsec, the blue triangle markers show the upper limit of DCO$^+$ emission in the inner 0.2 arcsec region of the disk, where we are limited by resolution. \textit{Bottom panel}: residuals obtained by subtracting the polynomial fit from the data are shown with black dots binned by 0.15 arcseconds. The blue solid line represents the gaussian that best fits the excess \cng{(see Table \ref{table:fit})}, with the shaded area showing the 1$\sigma$ fit uncertainty. The red triangle markers are the polarized light contrast in the Ks band \citepads[from][the scale is on the right side vertical axis]{garufi}.}
\label{fig:nodi}
\end{center}
\end{figure*}

\begin{figure*}[!ht]
\begin{center}
\includegraphics[keepaspectratio=true,width=8cm]{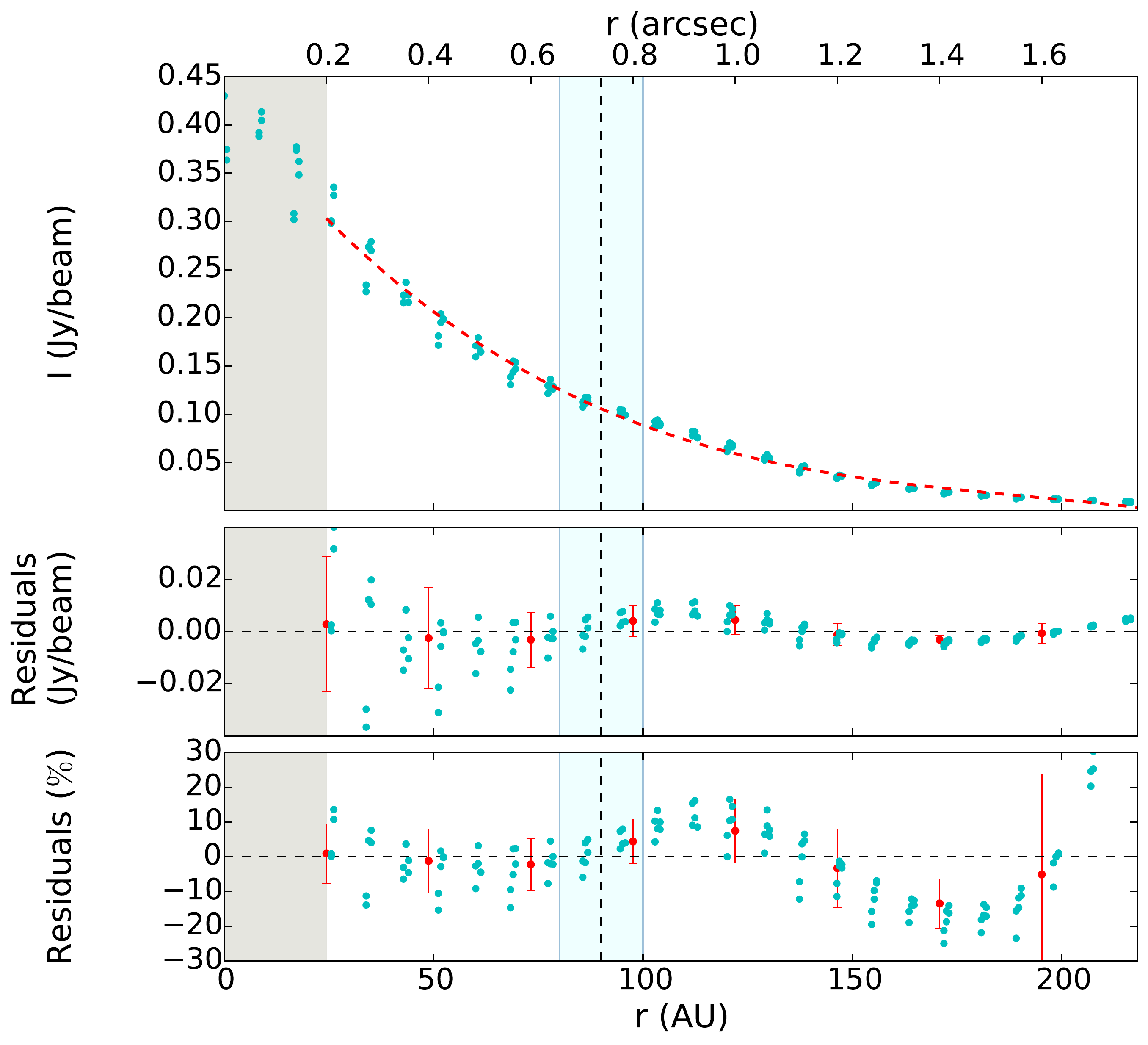}
\includegraphics[keepaspectratio=true,width=8cm]{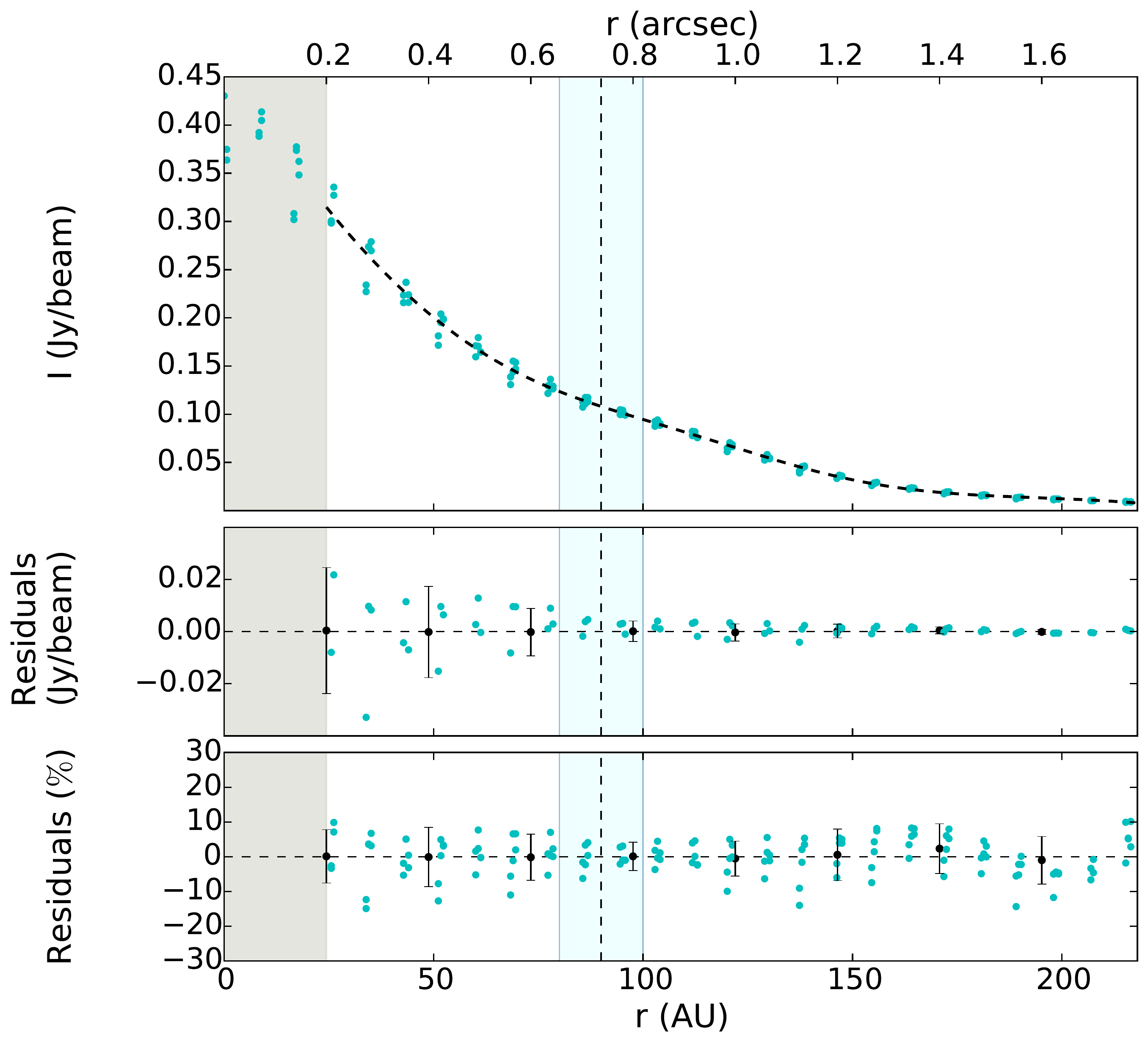}
\caption{{\it Left:} flux density across the disk major axis with a polynomial fit of degreee 3 (dashed line in the upper panel). In the second and third panel the absolute residuals and the percentage residuals respectively. {\it Right:} the same intensity profile fitted with a 3rd degreee polynomial plus a gaussian (see Table \ref{table:fit} for the best fit parameters).}
\label{fig:polii}
\end{center}
\end{figure*}

In Figure \ref{fig:nodi} we show the intensity profile of the image at 850 $\mu$m: 
as the disk is inclined by 44$\degree$ from the line of sight, the best angular resolution 
is reached using only the data along the projected disk major axis. We considered the pixels inside one beam across the major axis, every point corresponding to a pixel of 0.1~arcsec in the image (the points are therefore not all independent). The vertical spread is due to the shape and position angle of the synthesized beam, and to estimate the error when averaging on bins (Fig.~\ref{fig:nodi}, second panel) we weighted the points for the number of correlated pixels, i.e., $\sigma = \sqrt{\frac{\frac{1}{M}\Sigma_i (x_i - \mu )^2}{N/M -1}}$, where M is the number of correlated points and N is the number of averaged pixels. 
A simple analysis of the profiles reveals a bump in the emission between 80 and 150 AU: fitting a simple polynomial to outline a smooth profile does not produce an accurate fit (see Fig.~\ref{fig:polii}). The degree of the polynomial was chosen as the lowest degree that would provide a reasonable fit to the intensity profile.
To characterize the properties of this bump we fitted a combination of a 3rd degree polynomial plus a gaussian curve to our flux density profile (see Fig.~\ref{fig:polii}, right panel): subtracting the polynomial from the data leaves a gaussian-shaped residual (see Fig.\ref{fig:nodi}, second panel) centered at about 
(106~$\pm$~4)~AU, with a full width at half maximum (FWHM) of (71~$\pm$~18)~AU and a peak at (67$\pm$29)\% of the smooth polynomial profile. 
An estimate of the maximum spatial extent of the feature can be derived from deconvolving our best fit gaussian with the synthesized beam ($\sim$0.5$''$ in Band 7), resulting in an upper limit of $\sim$40~AU in FWHM. Note that these values are dependent on the choice of the pixel size of the image and the tolerance we use for the points on the major axis,  and thus are useful only for giving a rough estimate of the spatial scale of this unresolved emission excess. 
\cng{An independent analysis of this excess, obtained from modeling the visibilities directly, is shown in Section 4.} 

\begin{figure*}[ht!]
\begin{center}
\includegraphics[keepaspectratio=true,width=8cm]{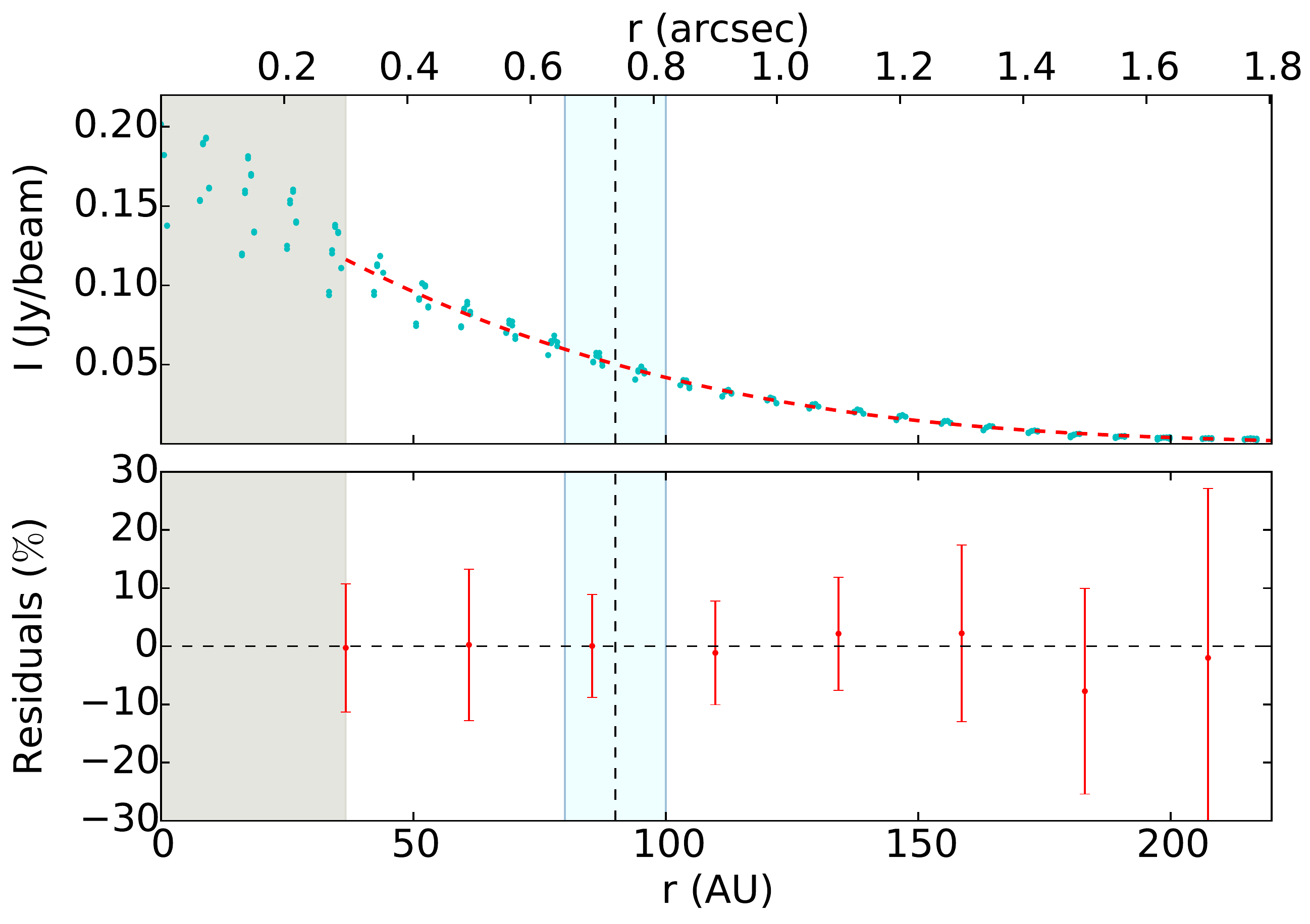}
\includegraphics[keepaspectratio=true,width=8cm]{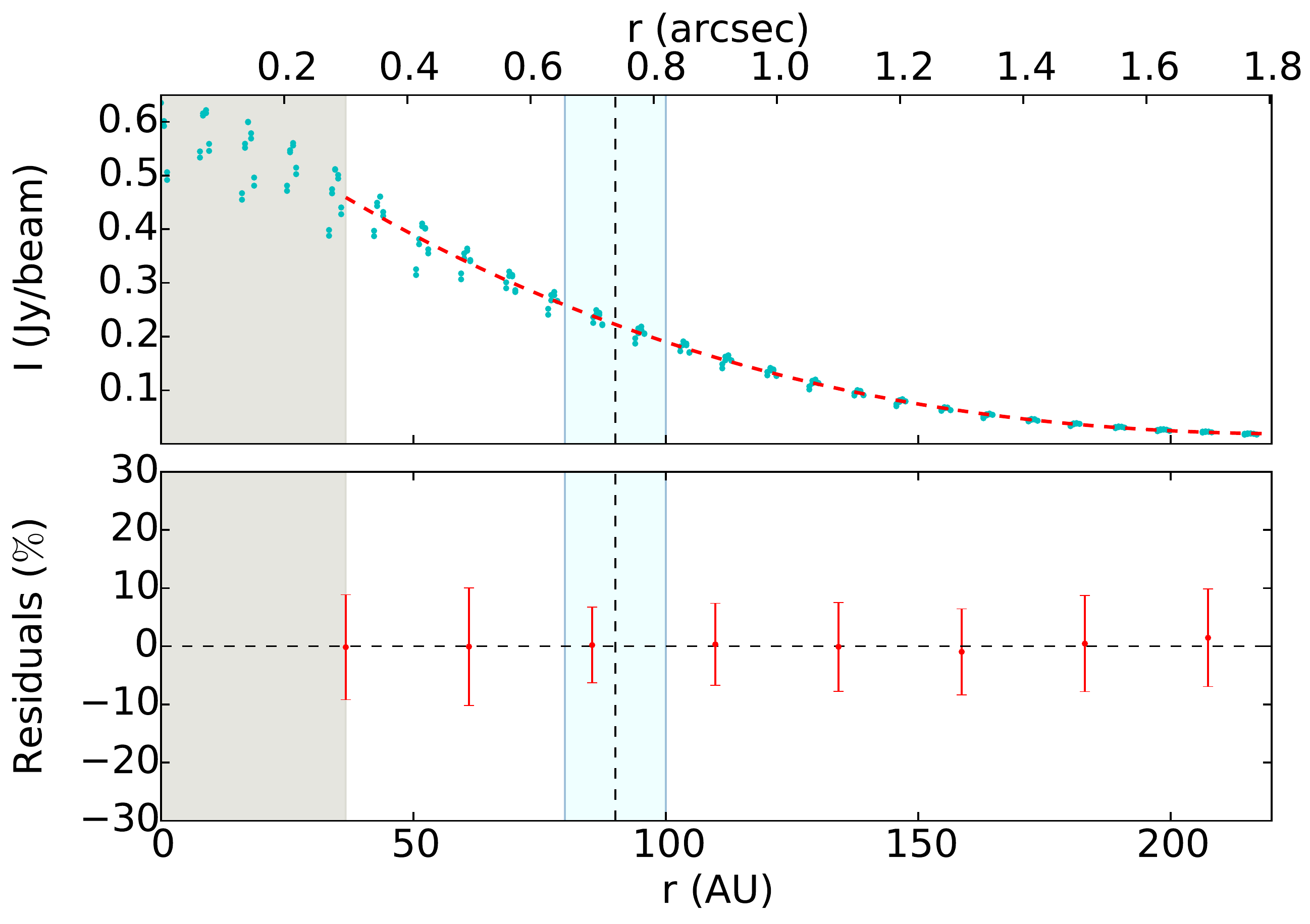}
\caption{{\it Left:} radial profile at 1.3~mm \cng{across the disk major axis}, with the \cng{dashed } curve resulting from the polynomial fit of the data. Bottom panel: absolute residuals with respect to the polynomial interpolation. The excess emission found at 850$\mu$m is not visible in Band~6. 
{\it Right}: radial profile of the image at 850~$\mu$m  restored with the same beam as Band~6 (0.74$''\times$0.60$''$, PA 77.3\degree).}
\label{fig:resb6}
\end{center}
\end{figure*}

The feature cannot be clearly identified in Band 6 intensity profiles: we find an indication of a faint excess in the radial profiles along the disk projected major axis, but its detection is dependent on 
the small variation of the position angle and inclination parameters, making its characterization unreliable. This is consistent with the lower angular resolution of the Band 6 SV data: if we image the Band 7 dataset with a restoring beam equal to Band 6 resolution, the feature is diluted and cannot be reliably separated from the smooth disk emission (see Figure \ref{fig:resb6}).

\begin{table}[h]
\resizebox{0.5\textwidth}{!}{
\begin{tabular}[]{l c c c c}
\toprule
\toprule
 & a$_0$ & a$_1$ & a$_2$ & a$_3$  \\
 Polynomial & (4.2$\pm$0.1)e-01 &(-5.6$\pm$0.3)e-03 &(2.7$\pm$0.2)e-05 &(-4.8$\pm$0.6)e-08   \\
 \midrule
 & a$_0$ & a$_1$ & a$_2$ & a$_3$  \\
\multirow{3}{1.5cm}{Polynomial +~Gaussian}&(4.6$\pm$0.2)e-01 &(-7.3$\pm$0.9)e-03&(4.0$\pm$0.7)e-05&(-7.4$\pm$1.6)e-8 \\
\cmidrule{2-5}
 & $\alpha$ [Jy/beam] & $\mu$ [AU] &$\sigma$ [AU] & \\
 & (3.4$\pm$1.5)e-02 & (1.06$\pm$0.04)e+02 & (3.0$\pm$0.8)e+01 & \\
\bottomrule
\end{tabular}}
\vspace{2mm}
\caption{Best fit parameters, with associated standard deviations, obtained from the least-square interpolation of the data along the major axis using a 3rd degree polynomial ($y=a_0+a_1x+a_2x^2+a_3x^3$) and a polynomial plus a gaussian curve ($y=a_0+a_1x+a_2x^2+a_3x^3+\alpha~exp(-(x-\mu)^2/2\sigma^2$), \cng{where $x$ is in AU and $y$ is in Jy/beam.}}
\label{table:fit}
\end{table}

\subsection{DCO$^+$ emission}

We extracted the DCO$^+$(J = 5$-$4) emission lines at 360.160 GHz from the ALMA Band 7 observations in order to compare the dust continuum radial profile with a potential molecular tracer of the CO snowline \citepads[see]{Mathews2013}. We used the CASA task ``clean'' with natural weighting to produce an integrated map of the DCO$^+$ emission in the velocity range 0.8-10~km/s, the resulting synthesized beam is $0.62\arcsec\times0.42\arcsec$. 
We find a ring-like structure, similar to that reported by \citetads[]{Mathews2013}, with a central radius of $\sim$110~AU and a total extent of the DCO$^+$ emission (detected at greater than 3$\sigma$) of 200~AU in radius. In Fig. \ref{fig:nodi} (top panel) we show the radial profile of the integrated DCO$^+$ emission along the disk projected major axis: we note a symmetry between the south-east and the north-west direction, both displaying a double peak at a distance of $\sim$60~AU and 140AU from the central star. We also point out that the minimum between the two peaks on both sides appears to fall at the position of the excess in the continuum emission at 850~$\mu$m ($\sim$110~AU). The signal to noise ratio of the DCO$^+$ image is very low, so this result is very tentative and should be verified with higher sensitivity observations.

\begin{figure}
\includegraphics[keepaspectratio=True,width=9cm]{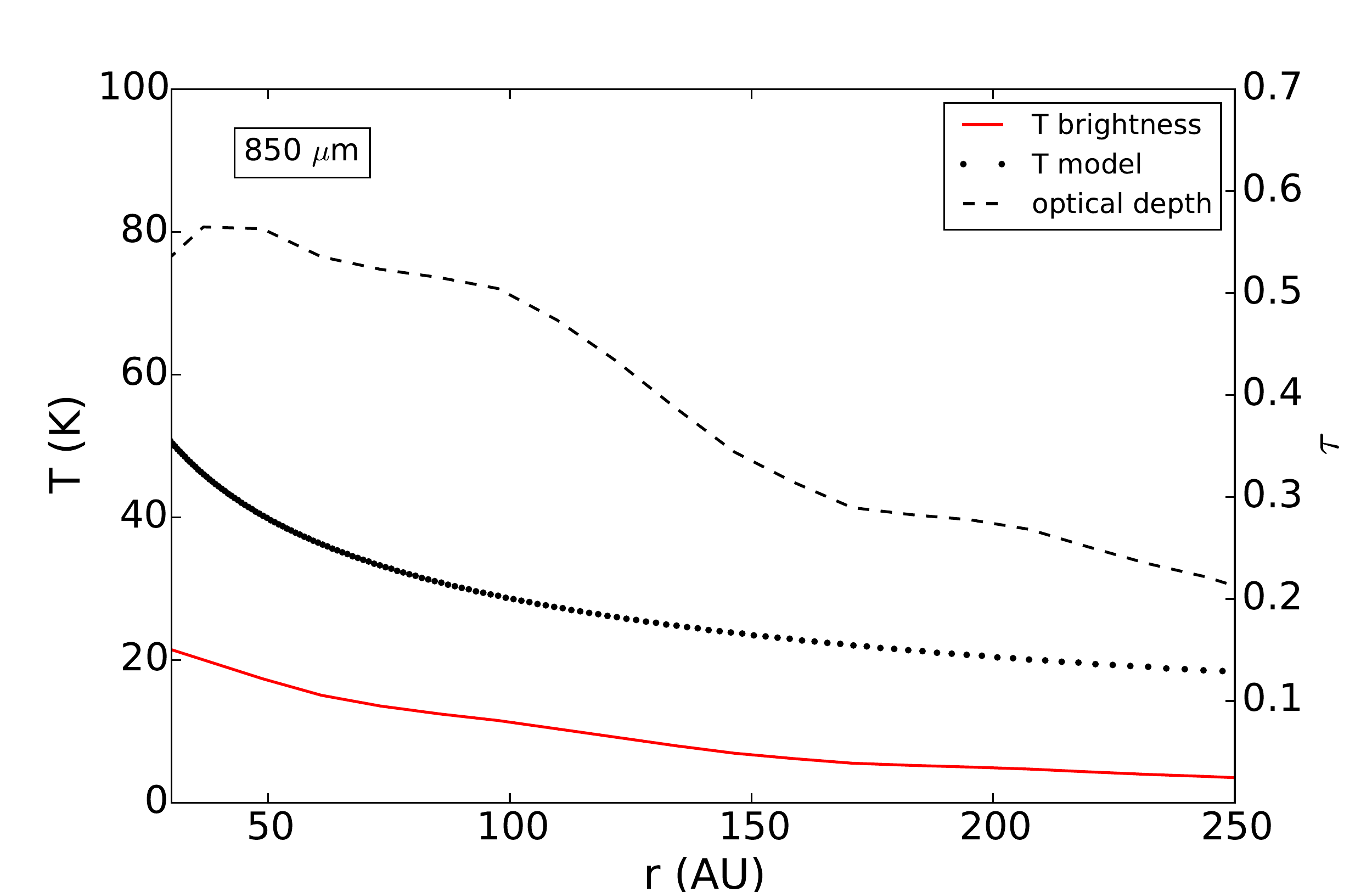}
\caption{Temperature profile of the best fit model (see Sect.\ref{sec:modello}) in black dots and brightness temperature from the observations at 850~$\mu$m (red solid line) plotted in function of the distance, starting from 30~AU to have a reliable estimate considering the resolution of the observations. On the right axis the optical depth from the comparison of the two temperatures, plotted as the dashed line.}
\label{fig:tau}
\end{figure}

\subsection{Spectral index profiles}
\label{sec:spidx}
\begin{figure*}[!ht]
 \centering
 \includegraphics[keepaspectratio=true,width=8cm]{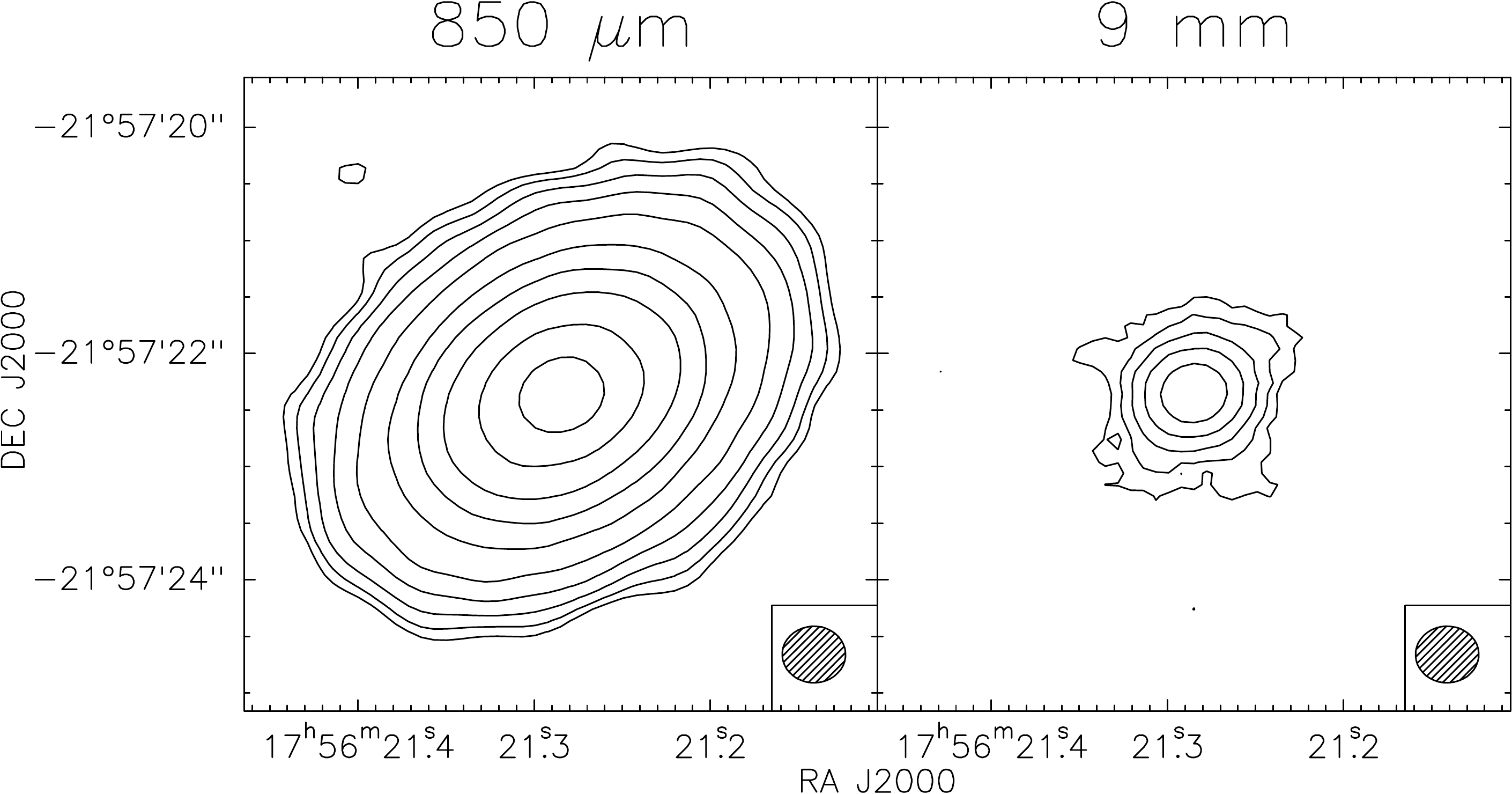}
 \hspace{0.6cm}
  \includegraphics[keepaspectratio=true,width=8cm]{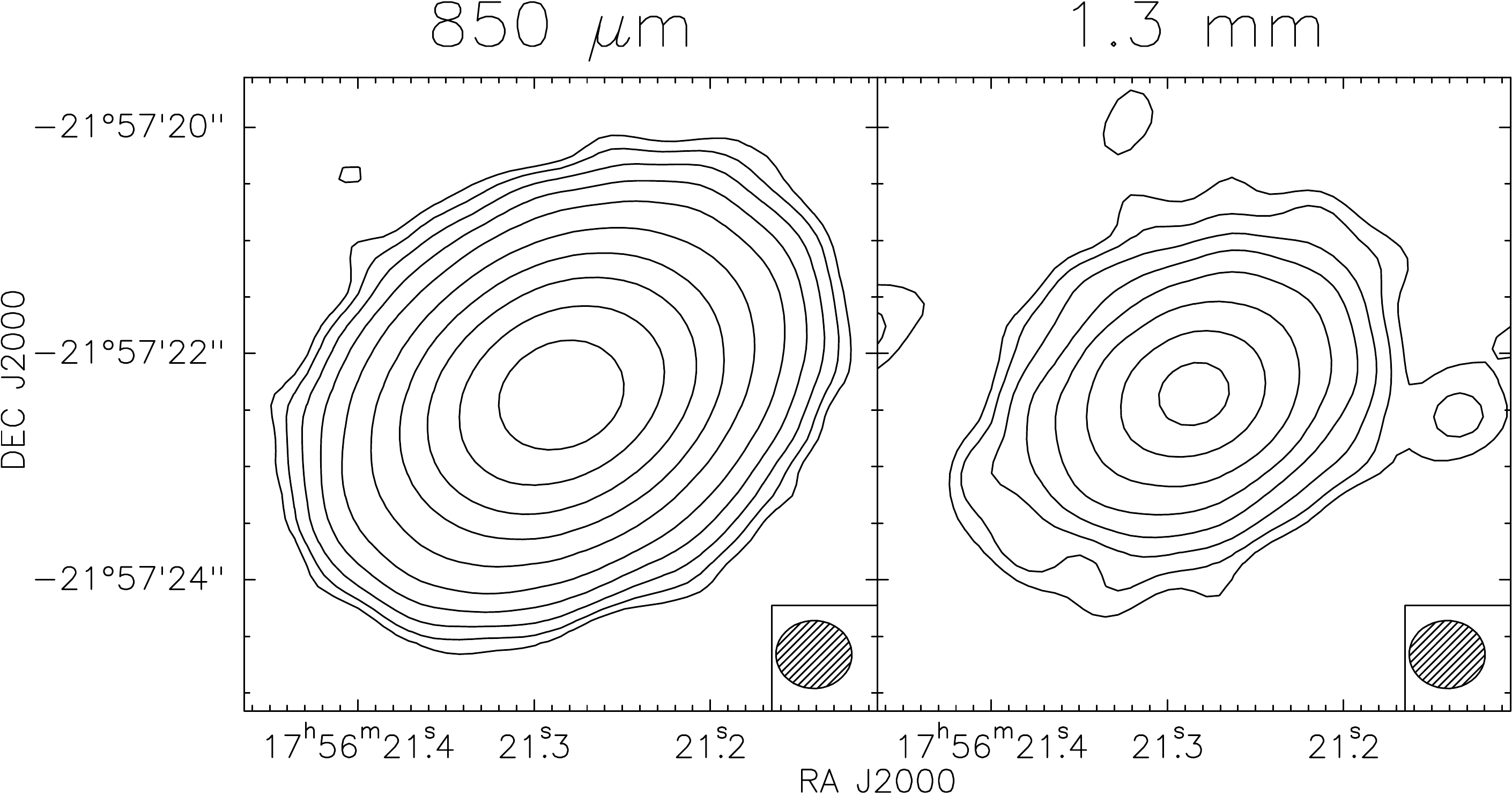}\\
\includegraphics[keepaspectratio=true,width=8.8cm]{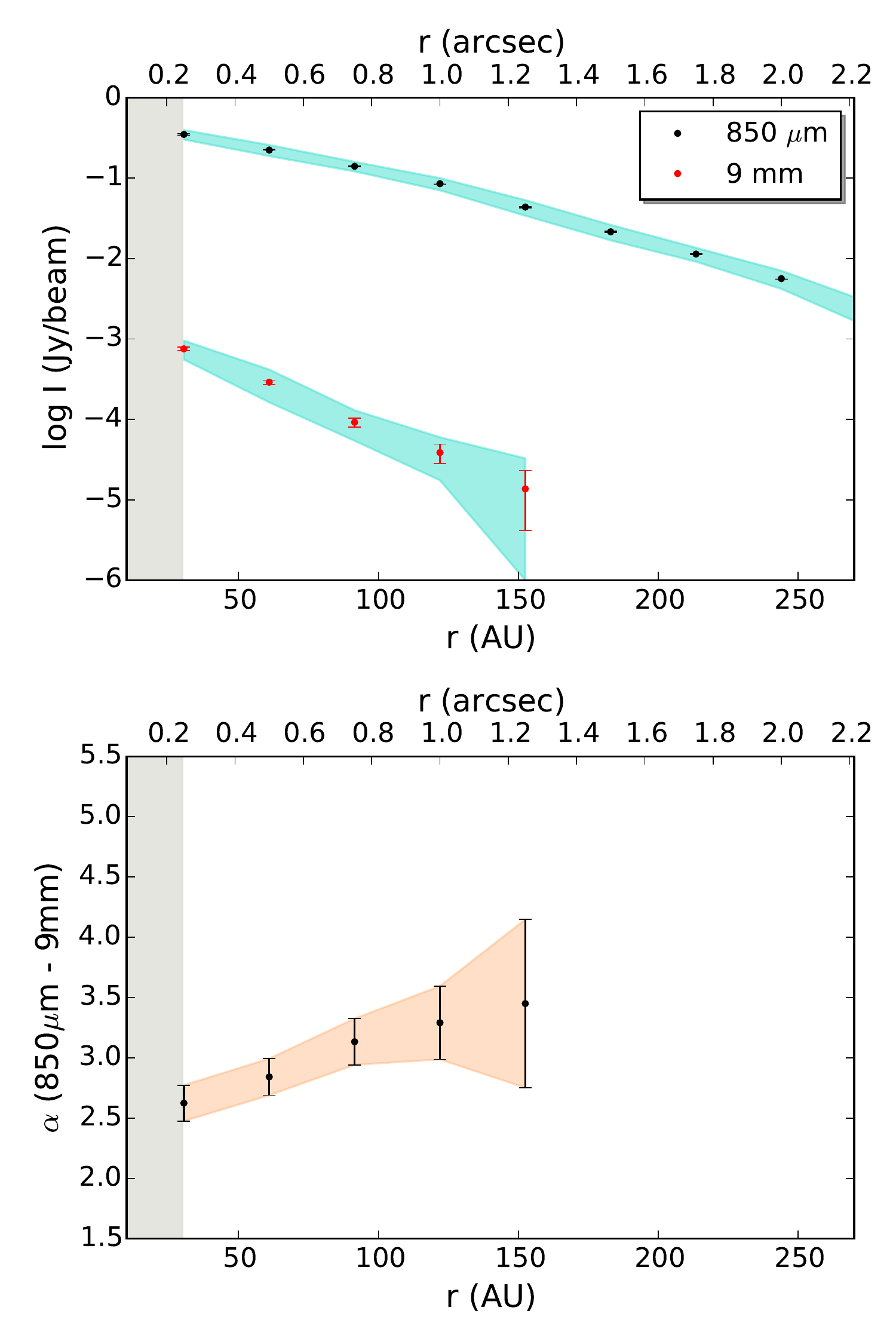}
 \includegraphics[keepaspectratio=true,width=8.8cm]{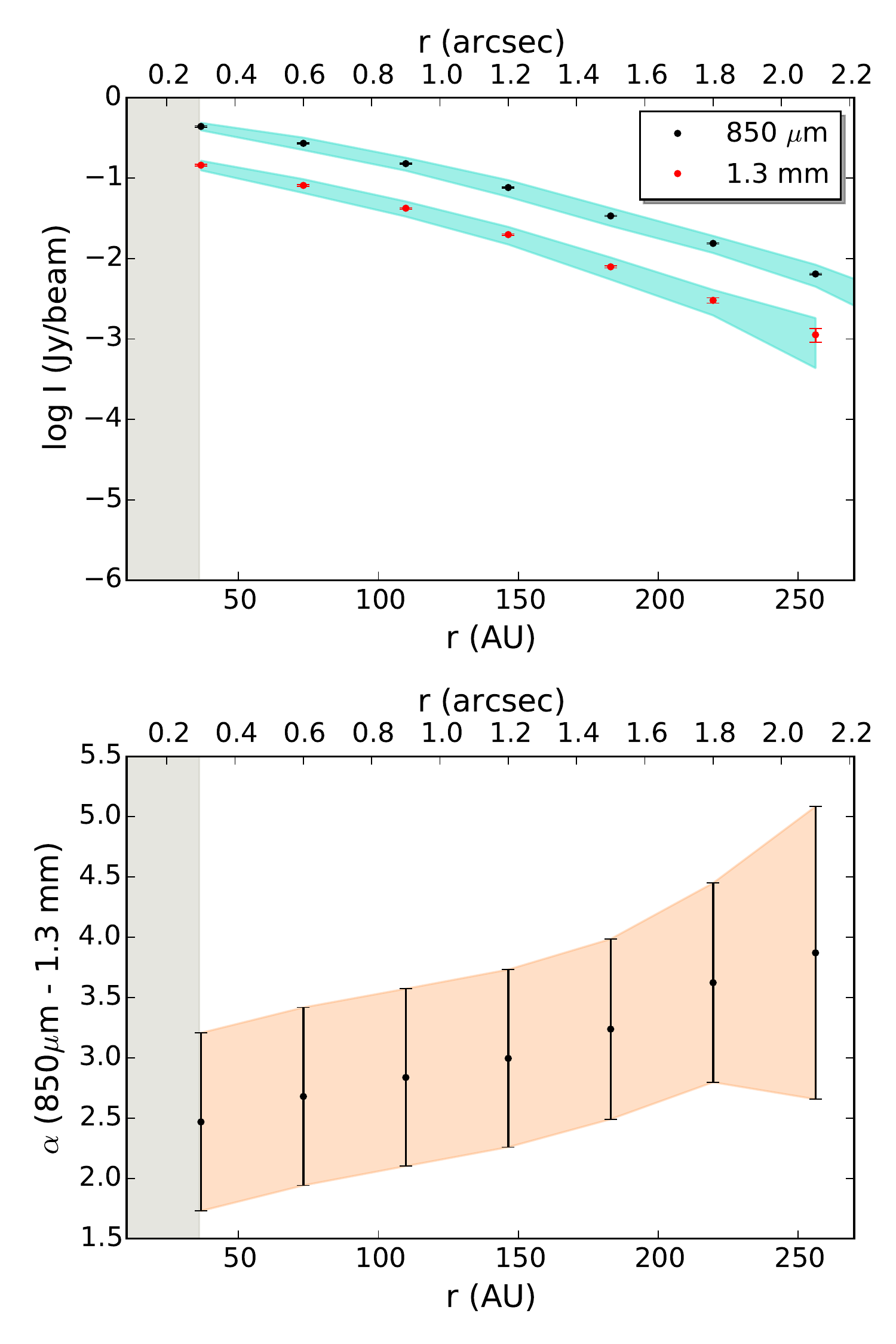}
  \caption{\textit{Left}: Flux density maps (top panel) at 850~$\mu$m and 9~mm, with contour levels at -3 (dashed), 3, 6, 12, 24, 48, 100, 200, 400, 800, 1600~$\sigma$. Disk surface brightness profiles (middle panel) used to compute the spectral index, plotted on a logarithmic scale. The shaded region shows the dispersion of the individual data points in the images, while the error bars show the uncertainty on the mean for each bin. Flux density spectral index (bottom panel) calculated between 850 $\mu$m and 9~mm as a function of radial distance. The profiles are averaged on ellipses of semi major axis $r$ (shown in arcsec and AU in the top and bottom axis, respectively). The shaded region indicates the uncertainty of the measurement, resulting from the noise of the deconvolved images and the systematic calibration uncertainty ; the low signal to noise ratio in the VLA map dominates the uncertainty on the spectral index determination beyond $\sim$100~AU. 
\textit{Right}: as above, but for the two ALMA datasets at 850~$ \mu$m and 1.3~mm.} 
 \label{fig:alphaevla}
\end{figure*}

The dust opacity at millimetre and submillimetre wavelengths is usually approximated by a power law $\kappa_\nu\propto\nu^\beta$ \citepads[e.g.,]{1983QJRAS..24..267H}. The emission properties depend on the details of the composition, geometry and size distribution of dust grains, most of which are very difficult to constrain. A general conclusion, however, is that in the conditions expected for dust in the densest regions of cores and disks, larger grain sizes correspond to lower values of the $\beta$ index \citepads[e.g.,]{1993Icar..106...20M, 1995A&A...296..797S, 2004ASPC..323..279N, 2006ApJ...636.1114D}. Even though connecting directly a value of $\beta$ to the detailed properties of the dust population is not possible, measurements of $\beta$ have been successfully used to infer the growth of dust in disks for many years \cng{\citepads[e.g.][]{1991ApJ...381..250B,2000ApJ...534L.101W,2001ApJ...554.1087T,2003A&A...403..323T,2006A&A...446..211R,2010A&A...512A..15R,kwon2015}}.

Previous measurements of the dust opacity power-law between 0.87 and 7~mm \citepads{isella} and between 1.3 and 7~mm \citepads{natta04} have already shown that grain growth occurs in the disk around HD163296. With the high resolution of ALMA it is now possible to extend these studies and constrain the radial behavior of the opacity spectral index $\beta$ \citepads[see, e.g.,][]{2012ApJ...760L..17P,Perez15}, not just its average value across the disk.  
We recall here that the emission from the disk midplane is generally optically thin and at these wavelengths the Rayleigh-Jeans regime is a good approximation. To verify these assumptions we compared the brightness temperature derived from our observations with the temperature profile of our best-fit model \cng{(see Sect. \ref{sec:modello})}, 
in order to estimate the optical depth~$\tau$ of the emission as $\tau \simeq -ln(1-T_b/T_{model})$. We found $\tau$ increasing towards the central regions, as expected, with values $<$0.6 for the emission at 850~$\mu$m, and $<$0.5 for the emission at 1.3~mm outside a inner region of $\sim$30~AU (see Fig.~\ref{fig:tau} for the 850~$\mu$m optical depth profile).
At longer wavelengths we estimate a lower optical depth, with values $\tau< 10^{-1}$ at both 8 and 10~mm. Hence the assumption of optically thin continuum emission is consistent with our data, and we expect a linear relation between flux density and dust opacity. 

The flux density emitted from a ring $dr$ at a given radius can be written as 
\begin{equation}
F_{\nu}(r) \propto \Sigma(r) \cdot B_\nu(T(r)) \cdot \nu^{\beta (r)},
\label{eq:flux}
\end{equation}
where $T(r)$ and $\Sigma$(r) are the midplane temperature and surface density at the distance $r$ from the star, and $B_\nu$ is the Planck function. So, in the Rayleigh-Jeans regime, 
we have: $F_{\nu}(r) \propto \Sigma(r) \cdot T(r) \cdot \nu^2 \cdot \nu^{\beta (r)}$. 
Producing matched images (same beam, pixel size and centered on the peak of the emission) at different wavelengths and measuring the ratio of the flux densities as a function of distance from the star allows the spectral index, $\alpha$ (where $F_\nu \propto \nu^\alpha$), to be determined.  Then, given the assumptions noted above, the power-law dependence of the dust opacity, $\beta$, can be derived as $\beta = \alpha - 2$, and does not depend on temperature or surface density. As discussed in Sec.~\ref{sec:sed} the observation at 8 and 10~mm includes gas emission from the stellar wind that needs to be subtracted to study the dust emissivity. The VLA maps used to compute the spectral index were produced subtracting a point source at the centre of the system with a flux density of 0.3~mJy from the calibrated visibilities (see Sec.~\ref{sec:sed}). 

In Figure \ref{fig:alphaevla}, left panel, we show the intensity maps at 850~$\mu$m and 9~mm with a circular beam of 0.5$^{\prime\prime}$, the corresponding averaged radial intensity profiles and the derived $\alpha$ profile. The intensity averaged values (and consequently the spectral index) are plotted as long as they stay above the 1$\sigma$ level. 
We see an increasing trend of the spectral index $\alpha$ from $\sim 2.5$ in the inner regions to $\sim 3.5$ at 150~AU. 
The profiles are displayed starting from and sampling every half resolution element of the images, corresponding to $\sim$30~AU; the large error associated with this spectral index profile is dominated by the limited signal to noise ratio of the VLA images.
With the same procedure we computed the spectral index between the  ALMA Band~6 and 7 images, with a matching circular beam of 0.6$^{\prime\prime}$, and obtained the profile shown in Figure 
\ref{fig:alphaevla}, right panel. In this case the profile  is sampled every 0.3$^{\prime\prime}$ ($\sim$40~AU). 
The large uncertainties, in spite of the high signal to noise of the ALMA images, is caused by the small wavelength leverage between the ALMA Band 6 and 7 observations. In propagating the uncertainty on $\alpha$ we used a 10\%\ calibration error for each flux, that in the case of $\alpha(850\mu m-1.3mm)$ represents a pessimistic estimate, as the two ALMA observations were carried out with the same phase calibrator and using Neptune as flux calibrator. 
Within the uncertainties 
the $\alpha$ profile seems to be consistent with the measurements between 850~$\mu$m and 9~mm. 
The spectral index remains below the value of 3 beyond 50 AU, corresponding to a $\beta$<1 under the assumptions mentioned above, and indicating the presence of grains that have grown to
at least 1~mm in size \citepads[using the dust opacity curves computed by][]{Testi2014}.  
We find no features in the $\alpha$ (and by implication, $\beta$) profiles across the CO snowline,
as would be expected for localized grain growth, but it should be noted that the resolution of the spectral index maps is limited by the lower resolution of Band 6 observations ($\sim$0.6$\arcsec$), which would not be sufficient to detect small scale variations of the emission (see Section \ref{subsec:excess.emission}). 
We return to the estimate of $\beta(r)$ in Sect.~\ref{sec:disc}.

\section{Modeling results}
\label{sec:modello}
In Section \ref{subsec:excess.emission} we have shown that the surface brightness radial profile for the continuum emission at 850~$\mu$m is compatible with having an excess peaked around 110\,AU, with deconvolved FWHM$\,\le 40$\,AU. In order to assess the robustness of this result we have performed a direct fit of the interferometric data (i.e., the visibilities) using the fitting scheme described by \citetads{Tazzari2015} to which we refer for the analysis details. \cng{In order to estimate the disk thermal emission at 850~$\mu$m, we use a classical two layer disk model \citepads{chiang97} with refinements by \citepads{dullemond2001} and a reduced disk flaring that adequately describes the observed far-infrared flux \citepads[from][]{tilling2012}. The resulting vertical scale height for the surface layer at $R>50$\,AU is $h/R\sim 0.08$.}\ Moreover, we assume a constant dust to gas mass ratio $\zeta = 0.01$ and the following gas surface density profile:
\begin{equation}
\label{eq:sigma.parametrization}
\Sigma_{\mathrm{g}}(R) = 
\Sigma_{0}\left(\frac{R}{R_{0}} \right)^{-\gamma}
\exp{\left[-\left(\frac{R}{R_{\mathrm{c}}} \right)^{2-\gamma} \right]}\,,
\end{equation}
where ${R_{0}=10\,}$AU is a fixed scale length and ${\Sigma_{0}}$, ${R_{c}}$ and $\gamma$ are free parameters to be fitted.
In order to compute the disk emission, the dust opacity is calculated using Mie theory \citepads[see][for details of the computation]{trotta2013} assuming the same dust composition throughout the disk, given by the following fractional abundances adapted from \citetads{1994ApJ...421..615P}: 5.4\%\ astronomical silicates, 20.6\%\ carbonaceaous material, 44\%\ water ice and 30\%\ vacuum. Furthermore, we assume a power-law grain size distribution $n(a)\propto a^{-q}$ for $a_\mathrm{min}\leq a\leq a_\mathrm{max}$, where $a$ is the grain radius. In order to model the fact that in the disk midplane we expect larger grains than on the surface \citepads{Testi2014}, we use different parametrization for the grain size distribution in these two regions: $q=3.5$, $a_\mathrm{min}=10\,$nm, $a_\mathrm{max}=100\mu$m in the surface and $q=3$, 
$a_\mathrm{min}=10\,$nm and $a_\mathrm{max}=0.8*(R/10~ AU)^{-1.025}$mm in the midplane, where this variable maximum grain size is chosen to reproduce the $\beta(R)$ profiles found in Sect.~\ref{sec:spidx}.
The modeling methodology is based on a Bayesian approach and employs an affine-invariant Markov Chain Monte Carlo (MCMC) ensemble sampler \citepads{2013PASP..125..306F} to explore the parameter space and find the best-fit models \citepads{Tazzari2015}.

\begin{figure}[ht!]
 \centering
 \includegraphics[keepaspectratio=true,width=0.5\textwidth]{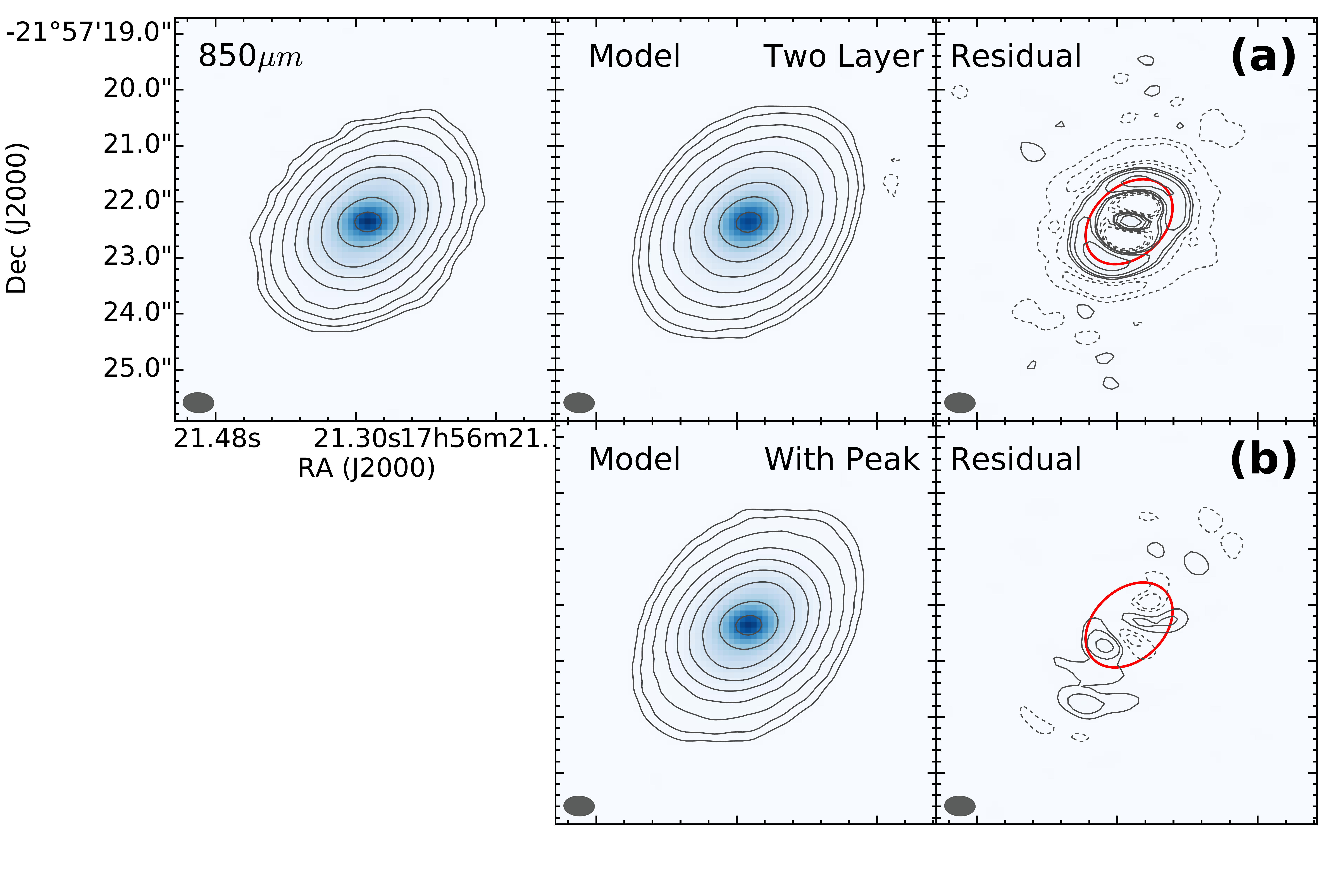}
 \caption{Continuum maps at $850\mu$m showing the results of the fits of the visibilities. The top left panel shows the observations, the two panels in the center the best-fit model, and the two panels on the right the residuals. The red solid curve represents the CO snowline at 90~AU. Panel~(a):~we use a classical two layer disk model which implements a monotonically decreasing surface brightness. Ring-shaped residuals are clearly visible. . Panel~(b):~we use the two layer disk model with an additional Gaussian peak. We fit the peak location ($R_{\mathrm{p}}=96\,$AU) and width FWHM$_{\mathrm{p}}=58\,$AU. The residuals are substantially reduced. The parameters used for the CLEAN are the same as discussed in Section~\ref{sec:obs} and the contour levels are the same as those used in Fig.~\ref{fig:cont} \cng{and Fig.~\ref{fig:alphaevla}}.}
 \label{fig:uvfit}
\end{figure}

In panel (a) of Figure \ref{fig:uvfit} we show the comparison between the observations at 850~$\mu m$ and the best-fit model (obtained running a MCMC with 500 chains) that corresponds to the following median values 
${\gamma=0.882\pm 0.002}$, ${\Sigma_{0}=(13.40\pm 0.03)\,}$g/cm$^2$, ${R_c=(118.7\pm 0.2)\,}$AU. 
We note that the model with median values also gives the minimum $\chi^2$, with $\chi^2_{\mathrm{red}}=$1.853). The ring-shaped residuals are clearly visible in the right plot of panel (a) and emphasize the fact that a simple two layer disk model with a monotonically decreasing surface density (and thus surface brightness) is not sufficient to completely account for the observed flux density profile. 

In order to assess whether the residuals can be explained by adding a simple ring-like peak or whether they need a more complicated treatment, we performed another fit with a modified version of the two layer model. This modified two layer disk model implements an additional ring-like structure in the 850~$\mu$m emission, with a disk surface brightness $I'(R)$ as follows:
\begin{equation}
I'(R) = I_{\mathrm{2L}}(R) +
I_p I_{\mathrm{2L}}(R_p)\exp\left[-\frac{(R-R_p)^2}{2\sigma_p^2} \right]\,,
\end{equation}
where $I_{\mathrm{2L}}(R)$ is the brightness computed by the classical two layer model, $R_p$ is the peak center, $\sigma_p$ is the peak width and $I_p$ is the peak intensity (in units of the brightness in the vicinity of the peak, namely $I_{\mathrm{2L}}(R_p)$).
This new model has therefore six free parameters: three of them for the two layer model ($\gamma,\ \Sigma_{0},\ R_{c}$) and another three to define the peak ($R_p,\ \sigma_p,\ I_p$). We perform the fit with the \citetads{Tazzari2015} modeling tool discussed above, and we obtain the results shown in panel (b) of Figure \ref{fig:uvfit}. The best-fit model ($\chi^2_{\mathrm{red}}=$1.830)  is described by ${\gamma=1.32\pm 0.01 }$, ${\Sigma_0=(24.6\pm 0.3)}$~g/cm$^2$, ${R_c=(111.3\pm 0.8)\,}$~AU, 
which correspond to a radial profile that is slightly steeper than the simple power-law model but has a similar cut-off radius. For the gaussian peak we find that is described by ${R_p=(96\pm 1)\,}$~AU, ${\sigma_p=(24.9\pm 0.5)\,}$~AU and ${I_p=128 \pm 20\%}$. 
This modified disk model is able to reproduce the observations with an extremely good agreement, as confirmed by the considerably smaller residuals (right plot of panel (b), Figure~\ref{fig:uvfit}).
The midplane temperature is computed at every radius according to the two-layer approximation; in Figure~\ref{fig:tau} (Section \ref{sec:spidx}) we show the temperature profile of this best-fit model. 

In conclusion, the peak inferred from the direct fit of the visibilities supports the evidence of a ring-like structure centered around 96\,AU with a  FWHM$=2\sqrt{2\ln 2}\cdot \sigma_p\approx (58 \pm 3)$\,AU, 
compatible with the upper limit resulting from the simple polynomial+gaussian fitting of the continuum surface brightness in Section \ref{subsec:excess.emission}.

\section{Discussion}
\label{sec:disc}

In Section~\ref{sec:res} we derived spatially resolved spectral index profiles for the dust emission from the HD~163296 protoplanetary disk and we identified and characterized an
unresolved excess 850\,$\mu$m emission centered at $\sim$110~AU. The presence of this feature has also been confirmed through a detailed modeling of the visibilities in Sect.~\ref{sec:modello}.
In this section we analyze these results and their possible implications for the growth of grains in the HD~163296 disk.

\subsection{$\beta(r)$ profiles and grain growth}

\begin{figure}[ht!]
 \centering
\includegraphics[keepaspectratio=true,width=8.8cm]{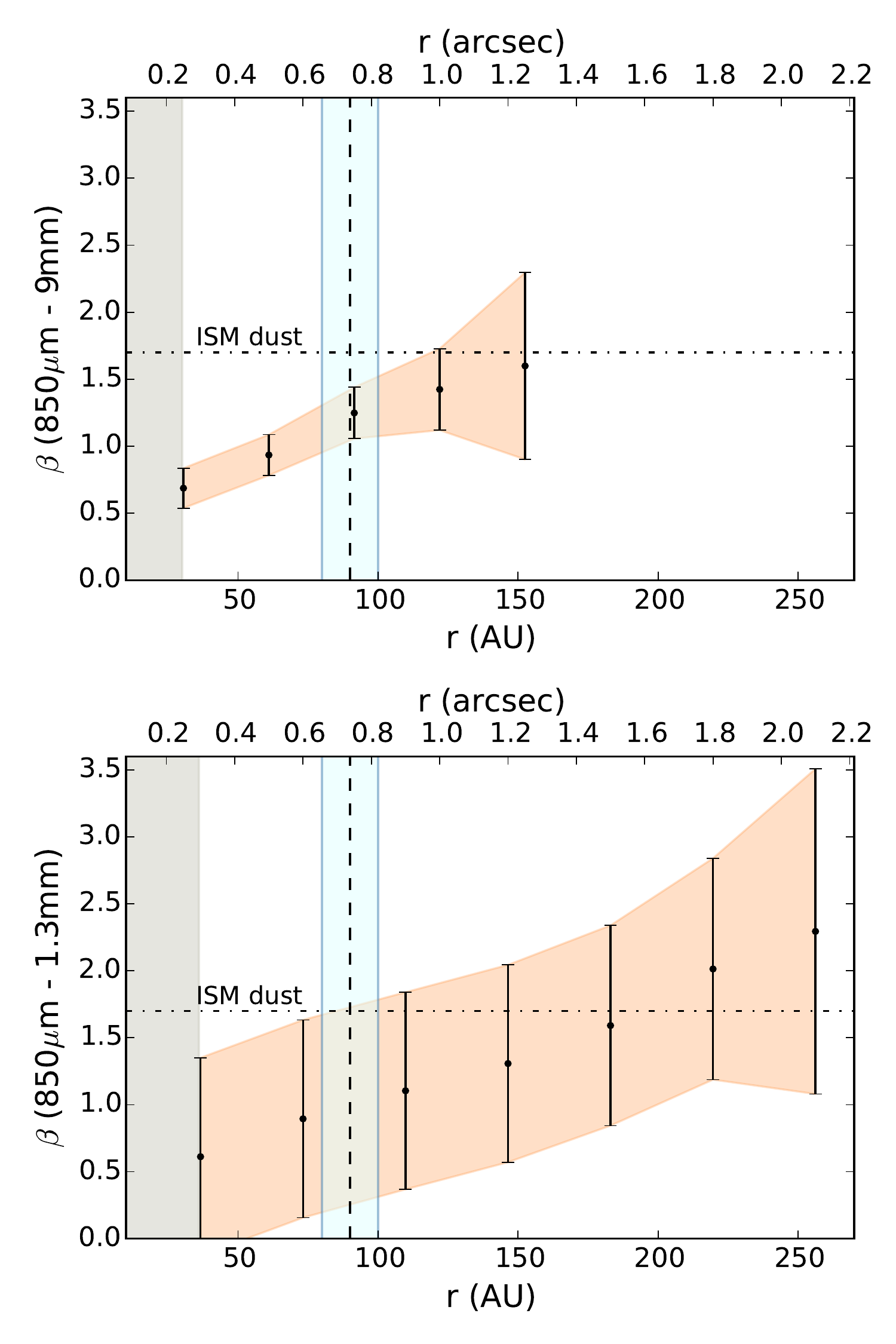}
 \caption{
Dust opacity power law index ($\beta$) profiles derived using Eq.~\ref{eq:beta}. $\beta(r)$ is computed between 850~$\mu$m and 9~mm in the top panel and between 850~$\mu$m and 1.3~mm in the bottom panel. The uncertainties are propagated from the surface brightness profiles presented in Fig.~\ref{fig:alphaevla}. The derived values of $\beta(r)$ in the top panel are dominated by the low signal to noise in the VLA image from $\sim$100~AU and are very uncertain beyond this radius. The CO snowline at 90$\pm$10 AU from \citetads{Qi2015} is pictured as the dashed vertical line.
}
 \label{fig:beta}
\end{figure}

In Section~\ref{sec:spidx} we derived the radial distribution of the spectral index as measured combining the ALMA~850\,$\mu$m image with the ALMA~1.3~mm or the VLA~10~mm images. 
Under the assumptions of optically thin emission and Rayleigh-Jeans regime, the spectral 
index profiles can directly be converted into opacity power law index profiles by subtracting a constant value of 2.0. Our modeling of the disk (Sect.~\ref{sec:modello}) and the comparison of the measured brightness temperature with the expected temperature profile from our model (Sect. \ref{sec:spidx}) confirm that the emission is optically thin throughout the disk, with the exception of the very inner region that is not resolved by the ALMA and VLA observations. On the other hand, the Rayleigh-Jeans approximation is not fully justified in the outer regions of the disk, especially for the ALMA Band~7 data. To estimate the value of the opacity power law index as a function of radius we thus used (see Eq.\ \ref{eq:flux}): 
\begin{equation}
\label{eq:beta}
\beta(r) = \left[ \log \large \left(\frac{\nu_1}{\nu_2}\right ) \right]^{-1} \left[ \log \left(\frac{F_{\nu_1}(r)}{F_{\nu_2}(r)}\right) - \log \left(\frac{B_{\nu_1}(T(r))}{B_{\nu_2}(T(r))}\right) \right]
\end{equation}
where $T(r)$ are the temperature profiles derived from our models. 
In Fig.~\ref{fig:beta} we show the profiles of $\beta(r)$ obtained from Eq.~\ref{eq:beta} using the intensity profiles shown in Fig.~\ref{fig:alphaevla}. 

Our analysis of the continuum emission at three different frequencies shows in the first place a decreasing spatial extent with increasing wavelength, confirming the presence of dust processing and radial transport in this disk, as already shown by the comparison between the size of the dust and gas disk by \citetads{itziar} and confirm and extend the results of  \citetads{natta04,2007prpl.conf..767N}, who showed the presence of large grains in the HD~163296 disk from integrated spectral indices and suggested a possible spectral index variation within the disk. 

The $\beta(r)$ profiles are qualitatively consistent with the results of similar analyses performed in other classical smooth disks \citepads[]{Guilloteau2011,Banzatti2011,2012ApJ...760L..17P,Perez15,trotta2013,menu2014,Tazzari2015}, with a significant variation of $\beta$ throughout the disk ($\Delta \beta>1$), indicating a maximum grain size $a_{max}$ decreasing with the distance from the star. The ALMA and VLA data provide for the first time the combination of signal to noise ratio, angular resolution and image fidelity to successfully perform an analysis in the image plane. Unfortunately, the ALMA Band~6 science verification data do not have the angular resolution to probe the dust properties in a localized region across the CO snowline. Similarly, the VLA data do not have a high enough signal to noise ratio at 100~AU and beyond to derive strong constraints in the outer disk.

The conclusion that we can draw from the $\beta(r)$ profiles is that there is convincing evidence for large grains inside the CO snowline, but at $\sim$70~AU resolution the profile appears to be smooth with no features. The data are consistent with significant grain growth throughout the inner 150-200~AU of the disk. As discussed by many authors
\citepads[see, e.g.,][and references therein]{Testi2014},
deriving a direct constraint on the level of grain growth from the $\beta$ values is not trivial, as it requires assumptions on the dust structure and composition, which cannot be constrained outside the Solar System. Nevertheless, a very broad range of reasonable assumptions on the grains properties
imply that $\beta$ values smaller than 1 in the millimeter to centimeter regime can only be produced by grains and pebbles larger than a millimeter in size, and can be significantly larger under reasonable assumptions for grain porosity \citepads{2004ASPC..323..279N}. As an example, if we adopt an educated guess for the grain composition based on the constraints from our own Solar System \citepads[e.g.,][]{1994ApJ...421..615P}, combined with a fraction of vacuum of $\sim$50\%, we derive 
maximum grain sizes as large as $\sim 1$~cm at the CO snowline and even exceeding $\sim 10$~cm in the inner $\sim$50~AU of the disk \citepads[see Fig.~4 of][]{Testi2014}.

\subsection{The nature of the 850~$\mu$m excess}

We found an excess emission at 850~$\mu$m located at about 105-115~AU from the star, with a full width along the disk major axis $\le$\,40~AU. 
This excess appears to be located very close to the CO snowline at 90 AU as measured by \cng{\citetads{Qi2015} } who resolved the N$_2$H$^+$ emission in this disk with ALMA. The N$_2$H$^+$ molecule is thought to be a robust tracer of CO condensation fronts, because of the strong correlation between its abundance and gas phase CO depletion \citepads[see also]{Qi2013}.

An analogous excess ring was found in the images of scattered light from HD 163296 taken with VLT/NACO \citepads{garufi}: polarized light images in Ks band displayed a $''$broken$''$ ring feature with an excess along the major axis between $\sim$0.5 and 1 arcsec, corresponding to 60 and 120 AU respectively (shown in Figure \ref{fig:nodi}, middle panel). The upper limit on the extent of the excess we found corresponds to a total radial extent of $\lesssim$40~AU, while the dimension of the resolved ring found in infrared polarized light by \citetads{garufi} measured 0.45$\arcsec$ in the east side and 0.6$\arcsec$ in the west side, corresponding to 60 AU and 73 AU respectively. 

The interpretation of the ring in Ks band polarized contrast given by \citetads{garufi} was the effect of self-shadowing of the disk created by a puffed-up inner region, with the outer disk emerging from the shadow at the location of the polarized emission. \citetads{garufi} could not 
exclude other effects, possibly related to the CO snowline, but could not reach a conclusion as the polarized infrared light is tracing the $\tau\sim$1 (at 2~$\mu$m) surface of the disk atmosphere, at much higher altitudes in the disk than the cold midplane where the CO~snowline (and the bulk of the disk material) is located. 

Our result provides an important contribution, as it shows that the excess is not purely a disk surface feature. The detection of the excess at the two different wavelengths,  tracing two different vertical regions of the disk, suggests the presence of a structure that concerns the whole vertical extent of the disk. 
At millimeter wavelengths we are probing the disk midplane and the emission is proportional to the surface density, the dust properties and the temperature profile (see Sect.~\ref{sec:spidx}). In principle any localized change of one (or more) of these properties can explain the excess emission that we find in our images. 

The lack of sensitivity or angular resolution in the VLA and ALMA Band~6 data do not allow us to probe the spectral index of the excess detected in Band~7. Future ALMA and/or VLA observations may allow us to probe the presence of large grains at the location of the excess emission.
The possibility of large grains at the snowline may also be connected with a local increase of the surface density (or temperature), which could also explain the observed excess. 

This is indeed expected from the simulations of grain growth across snowlines \citepads[e.g.,][]{RosJohansen2013}. We note that the effect of snowlines on grain growth is still poorly understood theoretically and much work is still needed. Recent simulations (Stammler, priv. comm.) 
show that, as grains maintain the (water) ice mantles across the CO snowline, this does not produce a discontinuity in the coagulation and fragmentation properties. The only effect would be a drop in the mass of solid particles inside the snowline as a fraction of the mantles is released in the gas. Such variation of the surface density distribution across the snowline may produce an effect similar to the one we observe in the brightness profile, and possibly also explain the effects on the disk surface observed in the near infrared. 

Another mechanism that has been proposed in order to explain emission rings near snowlines in disks is sintering \citepads{okuzumi2015}: this process brings icy grains to bond at temperatures close to the sublimation temperature. As a consequence, these aggregates can easily fragment by collisions close to the snowline leading to the accumulation of smaller fragments, which are less affected by radial drift. 

Clearly more theoretical work is needed before a detailed comparison of our observations with models can be done. 

An alternative explanation for this excess emission, not connected to the presence of the CO snowline, could be that particles are trapped by zonal flows \citepads[e.g]{dittrich2013} or by vortices \citepads[e.g.][]{klahr97}. Such a mechanism has been invoked to explain the presence of annular dust confinement in some transitional disks \citepads[e.g.,][]{2014ApJ...783L..13P}. Future higher angular resolution mm observations of the dust and gas will allow this possibility to be tested.

\section{Summary}
\label{sec:summary}

We have re-analyzed HD 163296 ALMA Science Verification data at 850 $\mu$m and 1.3~mm, and VLA data at 8 and 10~mm to study the radial behavior of dust properties in this disk. 
Our goal was to combine high resolution observations to derive the profile of the dust opacity spectral index, ultimately related to the size of grains, throughout the disk, and look for evidence of grain growth across the CO snowline. 

Our analysis shows in the first place more compact emission moving to longer wavelengths, confirming that dust processing and radial migration are taking place in this disk. 
A significant conclusion is that the dust opacity spectral index varies with radius and decreases towards the center to values $\leq$1, indicating the presence of large grains ($\geq$1~mm) in the inner regions of the disk (inside 100~AU). Our $\beta$(r) profiles are in agreement with those found in other resolved disks \citepads[e.g.,]{Guilloteau2011,2012ApJ...760L..17P,Perez15, Tazzari2015}. 

For this particular source, where a direct measurement of the location of the CO snowline is available, our analysis supports a scenario where the grains outside the snowline have not grown significantly, while the inner disk is populated by large grains. This general distribution would be consistent with an enhanced production of large grains at the CO snowline and subsequent transport to the inner regions; the alternative explanation of a smooth distribution of the grain sizes due to growth and transportation processes unrelated to the CO snowline is also consistent with the observed $\beta$(r) profile. \\

A second important finding is the evidence of an excess in the continuum emission at 850 $\mu$m near the location of the CO snowline and approximately at the same position of the excess in Ks band polarized light found by \citetads{garufi}. Our finding confirms that the infrared excess emission is not only related to a disk surface layer effect, but has more profound roots in the disk midplane, which is responsible for the 850~$\mu$m emission. The possible causes for this bump could be a local increase of the dust surface density due to dust trapping, e.g., caused by a local pressure maximum at the location of the snowline \citepads{armitage}, or by turbulent eddies that can retain grains in their interior \citepads{klahr97}. 

It is not clear if the dust at this location has a different opacity spectral index $\beta$ with respect to the bulk of the dust, since we lack the spatial resolution and/or sensitivity at 1.3~mm and $\sim$10~mm to clearly detect the excess emission. 

In order to probe conclusively if this excess is  a local change in the dust density and properties due to an effect of the CO snowline or another dust trapping process, ALMA high resolution and sensitivity observations at mm wavelengths are needed, as well as higher sensitivity VLA measurements. As a future development, with ALMA longer baselines we might be able to resolve the iceline of the most important volatile, H$_2$O, and investigate more extensively the role of snowlines in grain growth.

\begin{acknowledgements}
This paper makes use of the following ALMA data: ADS/JAO.ALMA\#2011.0.000010.SV. ALMA is a partnership of ESO (representing its member states), NSF (USA) and NINS (Japan), together with NRC (Canada), NSC and ASIAA (Taiwan), and KASI (Republic of Korea), in cooperation with the Republic of Chile. The Joint ALMA Observatory is operated by ESO, AUI/NRAO and NAOJ. 
We are grateful to Antonio Garufi for many discussions and for sharing his infrared polarization images. We thank Sebastian Stammler for insightful discussions on the effect of the CO snowline on dust and for showing us the results of his simulations in advance of publication. We thank the anonymous referee for the helpful comments. Part of this research was carried out at the Jet Propulsion Laboratory, California Institute of Technology, under a contract with the National Aeronautics and Space Administration. The fits have been carried out on the computing facilities of the Computational Center for Particle and Astrophysics (C2PAP) as part of the approved project ``Dust evolution in protoplanetary disks''. MT and LT are grateful for the experienced support by F. Beaujean (C2PAP). This work was partly supported by the Italian Ministero dell\'\,Istruzione, Universit\`a e Ricerca through the grant Progetti Premiali 2012 -- iALMA (CUP C52I13000140001).
\end{acknowledgements}

\bibliography{biblio}
\bibliographystyle{aa} 

\end{document}